%% file: draft.tex
\documentclass[longauth,traditabstract]{aa} 
\usepackage{graphicx}
\usepackage{lscape}
\usepackage{txfonts}
\usepackage{natbib}
\usepackage{url}
\usepackage[T1]{fontenc}

\usepackage[bookmarks=false]{hyperref}

\newcommand{\FunitsA}{10$^{-16}$ erg~s$^{-1}$~cm$^{-2}$~arcsec$^{-2}$}

\newcommand{\nodata}{ ~$\cdots$~ }

\newcommand{\fluxA}{erg\,s$^{-1}$\,cm$^{-2}$\,\AA$^{-1}$}

\newcommand{\mz}{{\small $\mathcal{M}$-Z}}
\newcommand{\Sz}{{\small {\large $\Sigma$}-Z}}

\DeclareRobustCommand{\ion}[2]{%
\relax\ifmmode
\ifx\testbx\f@series
{\mathbf{#1\,\mathsc{#2}}}\else
{\mathrm{#1\,\mathsc{#2}}}\fi
\else\textup{#1\,{\mdseries\textsc{#2}}}%
\fi}

\newcommand{\HII}{\ion{H}{ii}~}

\begin{document}
   \title{The Mass-Metallicity relation explored with CALIFA:}
   \subtitle{I. Is there a dependence on the star formation rate?}

   \author{
     S.\,F. S\'anchez\inst{\ref{iaa},\ref{caha},\ref{prague}}
     \and
     F.\,F. Rosales-Ortega\inst{\ref{uam},\ref{inaoe}}
     \and
     B. Jungwiert\inst{\ref{prague}}
     \and
     J. Iglesias-P\'aramo\inst{\ref{iaa},\ref{caha}}
     \and
     J.\,M. V\'\i lchez\inst{\ref{iaa}}
     \and
     R.\,A. Marino\inst{\ref{ucm}}
     \and
     C.\,J.\, Walcher\inst{\ref{aip}}
     \and
     B.\, Husemann\inst{\ref{aip}}
     \and
     D. Mast\inst{\ref{caha},\ref{iaa}}
     \and
     A. Monreal-Ibero\inst{\ref{iaa}}
     \and
     R. Cid Fernandes\inst{\ref{flora}}
     \and
     E. P\'erez \inst{\ref{iaa}}
     \and
     R. Gonz\'alez Delgado\inst{\ref{iaa}}
     \and
     R. Garc\'\i a-Benito \inst{\ref{iaa}}
     \and
     L. Galbany\inst{\ref{centra}}
     \and
     G. van de Ven \inst{\ref{mpia}}
     \and
     K. Jahnke \inst{\ref{mpia}}
     \and
     H. Flores \inst{\ref{odp}}
     \and
     J. Bland-Hawthorn \inst{\ref{sydney}}
     \and
     A. R. L\'{o}pez-S\'{a}nchez \inst{\ref{aao}}
     \and 
     V. Stanishev \inst{\ref{centra}}
     \and
     D.\,Miralles-Caballero\inst{\ref{uam}}
     \and
     A.\,I.\, D\'\i az\inst{\ref{uam}}
     \and
     P.\, S\'anchez-Blazquez\inst{\ref{uam}}
     \and
     M. Moll\'a\inst{\ref{ciemat}}
     \and
      A. Gallazzi\inst{\ref{dark}}
     \and
      P. Papaderos \inst{\ref{porto}}
     \and
      J. M. Gomes\inst{\ref{porto}}
     \and
      N. Gruel\inst{\ref{sheffield}}
     \and
      I. P\'erez\inst{\ref{ugr}}
     \and
      T. Ruiz-Lara\inst{\ref{ugr}}
     \and
      E. Florido\inst{\ref{ugr}}
     \and
      A. de Lorenzo-C\'aceres\inst{\ref{lalag}}
     \and
      J. Mendez-Abreu \inst{\ref{iac},\ref{lalag}}
     \and
      C. Kehrig\inst{\ref{iaa}}
     \and
      M.M. Roth\inst{\ref{aip}}
     \and
      B. Ziegler\inst{\ref{viena}}
     \and
      J. Alves\inst{\ref{viena}}
     \and
     L.\, Wisotzki\inst{\ref{aip}}
     \and
     D.\, Kupko\inst{\ref{aip}}
     \and
     A. Quirrenbach\inst{\ref{land}}
     \and
     D. Bomans\inst{\ref{bochum}}
     \and     
     { The CALIFA collaboration}
          }

   \institute{
        \label{iaa}Instituto de Astrof\'{\i}sica de Andaluc\'{\i}a (CSIC), Glorieta de la Astronom\'\i a s/n, Aptdo. 3004, E18080-Granada, Spain\\ \email{sanchez@iaa.es}.
        \and
        \label{caha}Centro Astron\'omico Hispano Alem\'an, Calar Alto, (CSIC-MPG),
        C/Jes\'{u}s Durb\'{a}n Rem\'{o}n 2-2, E-04004 Almer\'{\i}a, Spain.
\and
\label{prague}Astronomical Institute, Academy of Sciences of the Czech Republic, Bo\v{c}n\'{i} II 1401/1a, CZ-141 00 Prague, Czech Republic.
       \and
       \label{uam}Departamento de F\'isica Te\'orica, Universidad Aut\'onoma de Madrid, 28049 Madrid, Spain.
       \and
       \label{inaoe} Instituto Nacional de Astrof{\'i}sica, {\'O}ptica y Electr{\'o}nica, Luis E. Erro 1, 72840 Tonantzintla, Puebla, Mexico
       \and
\label{ucm}CEI Campus Moncloa, UCM-UPM, Departamento de Astrof\'{i}sica y CC$.$ de la Atm\'{o}sfera, Facultad de CC$.$ F\'{i}sicas, Universidad Complutense de Madrid, Avda.\,Complutense s/n, 28040 Madrid, Spain.
\and
        \label{aip}Leibniz-Institut f\"ur Astrophysik Potsdam (AIP), An der Sternwarte 16, D-14482 Potsdam, Germany.
\and
\label{flora}Departamento de F\'{\i}sica, Universidade Federal de Santa Catarina, P.O. Box 476, 88040-900, Florian\'opolis, SC, Brazil
\and
\label{centra}CENTRA - Instituto Superior Tecnico, Av. Rovisco Pais, 1, 1049-001 Lisbon, Portugal. 
 \and
\label{mpia}Max-Planck-Institut f\"ur Astronomie, Heidelberg, Germany. 
\and
\label{odp}Laboratoire Galaxies Etoiles Physique et Instrumentation, Observatoire de Paris, 5 place Jules Janssen, 92195 Meudon, France
\and
        \label{sydney}Sydney Institute for Astronomy, School of Physics A28, University of Sydney, NSW 2006, Australia.
        \and
        \label{aao}Australian Astronomical Observatory, PO BOX 296, Epping, NSW 1710, Australia.
        \and
        \label{ciemat}Departamento de Investigaci\'on B\'asica, CIEMAT, Avda. Complutense 40 E-28040 Madrid, Spain.
 \and
\label{dark}Dark Cosmology Centre, Niels Bohr Institute, University of Copenhagen, Juliane Maries Vej 30, DK-2100 Copenhagen, Denmark.
\and
\label{porto}Centro de Astrof\'\i sica and Faculdade de Ciencias,
Universidade do Porto, Rua das Estrelas, 4150-762 Porto, Portugal.
\and
\label{sheffield}University of Sheffield, Department of Physics and Astronomy, Hicks Building, Hounsfield Road, Sheffield, S3 7RH, United Kingdom.
\and
\label{ugr}Dpto. de F\'\i sica Te\'orica y del Cosmos, University of Granada, Facultad de Ciencias (Edificio Mecenas), 18071 Granada, Spain 
\and
\label{lalag}Depto. Astrof\'\i sica, Universidad de La Laguna (ULL), E-38206 La Laguna, Tenerife, Spain
\and
\label{iac}Instituto de Astrof\'\i sica de Canarias (IAC), E-38205 La Laguna, Tenerife, Spain 
\and
\label{viena}University of Vienna, T\"urkenschanzstrasse 17, 1180 Vienna, Austria.
\and
\label{land} Landessternwarte, Zentrum für Astronomie der Universität Heidelberg,
Königstuhl 12, D-69117 Heidelberg, Germany
\and
\label{bochum}  Astronomical Institute of the Ruhr-University Bochum Universitaetsstr. 150, 44801 Bochum, Germany.
     \thanks{Based on observations collected at the Centro Astron\'omico
      Hispano Alem\'an (CAHA) at Calar Alto, operated jointly by the Max-Planck
      Institut f\"ur Astronomie and the Instituto de Astrof\'{\i}sica de Andaluc\'{\i}a (CSIC).}
              }

   \date{Received ----- ; accepted ---- }

 
\abstract{ 
We present the results on the study of the global and local
  \mz\ relation based on the first data available from the CALIFA
  survey (150 galaxies). This survey provides integral field
  spectroscopy of the complete optical extent of each galaxy (up to
  2-3 effective radii), with enough resolution to separate
  individual \HII regions and/or aggregations. Nearly $\sim$3000
  individual \HII\ regions have been detected. The spectra
  cover the wavelength range between [OII]3727 and
  [SII]6731, with a sufficient signal-to-noise to derive the oxygen
  abundance and star-formation rate associated with each region. In
  addition, we have computed the integrated and spatially resolved
  stellar masses (and surface densities), based on SDSS photometric data.
   We explore the relations between the stellar mass, oxygen abundance
   and star-formation rate using this dataset.

We derive a tight relation between the integrated stellar mass and the
gas-phase abundance, with a dispersion smaller than the one already
reported in the literature ($\sigma_{\Delta{\rm log(O/H)}}=$0.07
dex). Indeed, this dispersion is only slightly larger than the typical
error derived for our oxygen abundances. However, we do not find
any secondary relation with the star-formation rate, other than the
one induced due to the primary relation of this quantity with the stellar
mass. The analysis for our sample of $\sim$3000 individual
\HII\ regions confirm (i) the existence of a local
mass-metallicity relation and (ii) the lack of a secondary relation
with the star-formation rate. The same analysis is done for the specific
star-formation rate, with similar results.

Our results agree with the scenario in which gas recycling in
galaxies, both locally and globally, is much faster than other typical
timescales, like that of gas accretion by inflow and/or metal loss due to
outflows. In essence, late-type/disk dominated galaxies seem to be in
a quasi-steady situation, with a behavior similar to the one expected
from an instantaneous recycling/closed-box model.

}

   \maketitle


\section{Introduction}

Metals form in stars as a by-product of the thermonuclear reactions
that are the central engine of stellar activity. Once they have completed their
life cycle, stars eject metals into the interstellar medium, polluting
the gas, which is the fuel for the new generation of stars. Therefore,
the star-formation rate (SFR), the stellar mass, the metal content and
the overall star-formation history of galaxies are strongly
interconnected quantities. A fundamental open question in our
understanding of galaxy evolution is based on the details of these
interconnections (are they local or global? are both intristic or a
consequence of the evolution?), and their dependence on other
properties of galaxies (are they affected by the environment and how?
how does the merging history of the small interactions affect them?).

The existence of a strong correlation between stellar mass and
gas-phase metallicity in galaxies is well known 
\citep{leque79,skill92}.  These parameters are two of the most fundamental
physical properties of galaxies, both directly related to the process
of galaxy evolution.  The mass-metallicity (\mz) relation is
consistent with more massive galaxies being more metal-enriched. It
was confirmed observationally by \citet[][ hereafter T04]{tremonti04}, who found a tight correlation spanning over 3
orders of magnitude in mass and a factor of 10 in metallicity, using a
large sample of star-forming galaxies up to z\,$\sim$\,0.1 from the
Sloan Digital Sky Survey (SDSS). The \mz\ relation appears to be
independent of large-scale \citep{Mouhcine:2007p4175} and local
environment \citep{hugh12}, although it has been observed that in
high density environments metallicities are higher than expected \citep[e.g.][]{mate07,petro12}, and in lower density environments satellities have higher metallicities than central galaxies \citep[e.g.][]{pasq10,pasq12}. Finally, the relation has been confirmed at all accessible
redshifts \citep[e.g.][]{sava05,erb06,maio08}.

Considerable work has been devoted to understanding the physical
mechanisms underlying the \mz\ relation. The proposed scenarios to
explain its origin can be broadly categorized as: 1) the loss of
enriched gas by outflows \citep[T04;][]{koba07}; 2) the
accretion of pristine gas by inflows \citep{fila08}; 3)
variations of the initial mass function with galaxy mass
\citep{kopp07}; 4) selective star formation efficiency or
{\em downsizing} \citep{broo07,elli08,calu09,vale09}; or a combination of them. 

Despite the {\it local} nature of the star-formation processes, 
it has only recently become possible to analyze the \mz\ relation
for spatially resolved, external galaxies. Early precedents
with more limited data include \citet{edmu84} and \citet{VilaCostas:1992p322},
who noticed a correlation between mass surface density and gas
metallicity in a number of galaxies. \citet{moran12} recently reported
a correlation between the local stellar mass density and the
metallicity valid across all the galaxies in their sample. This result
was derived analyzing the individual \HII\ regions of a sample of 174
star-forming galaxies, based on slit spectroscopy. We 
independently confirmed this relation \citep{rosales12}, using a statistically
complete sample of $\sim$2500 \HII\ regions extracted from a sample of
38 galaxies, using integral field spectroscopy (IFS), described in
\citet{sanchez12b}. 

This local relation is a scaled version of the \mz\ relation. This new relation is explained as a simple effect of
the {\em inside-out} mass/metallicity growth that dominates the
secular evolution of late-type galaxies \citep[e.g.][]{matt89,bois99},
combined with the fact that more massive regions form stars faster
(i.e. at higher SFRs), thus earlier in cosmological times, which can
be considered a local {\em downsizing} effect, similar to the one
observed in individual galaxies \cite[e.g.][]{per-gon08}. This
explanation does not require a strong effect of inflows/outflows in
shaping the \mz\ relation, which can be naturally explained by secular
evolution processes.

Recently, evidence has been reported for a dependence of the
\mz\ relation on the SFR. Different authors found that these three
quantities define either a surface or a plane in the corresponding 3D
space, i.e., the so-called Fundamental Mass-Metallicity relation (FMR)
and/or \mz-Fundamental Plane, depending on the authors
\citep{lara10a,mann10,yate12}. Regardless of the actual shape of this
relation \citep{lara12}, its existence is still controversial. The
described relationship implies that (i) for the same mass, galaxies with
stronger star-formation rates, have lower metallicities, and (ii) for
low-mass galaxies, the dependence on the SFR is stronger. Metallicity
is a parameter that depends on (i) the star-formation (which enriches
galaxies), (ii) inflows (which dilute and reduce the metallicity), and
(iii) outflows (which eject metals out of the galaxy). Thus, the
reported relation imposes restrictions to the ratio between the
chemical enrichment, produced by the SFR, and dynamical timescales
which regulates the dilution due to inflows \cite[e.g.][]{quil08}. On
the other hand, it also restricts the dependence between the amount of
metals ejected by an outflow and the strength of the SFR \cite[][ and
  references therein]{mann10}. In a stable situation, described by a
simple instantaneous recycling model, no dependence on the SFR is
expected.

So far, all the studies reporting a dependence between the
\mz\ relation and the SFR are based on single aperture spectroscopic
data, like the one provided by the SDSS survey \citep{york00}. In spite 
of the impressive dataset provided by these surveys, with tens or
hundreds of thousands of individual measurements, they present some
well-known drawbacks: (i) single aperture spectroscopic data have a
fundamental aperture bias, which restricts the derived parameters to
different scale-lengths of the galaxies at different redshifts \citep[e.g.][]{ellis05}. This
is a strong limitation for these studies, considering that both the
oxygen abundance and the star-formation rate shows strong gradients
within galaxies \citep[e.g.][]{sanchez12b}. Therefore, the derived
parameters may not be representative of the integrated (or
characteristic) ones; (ii) The fact that most of these surveys sample
galaxies at a wide range of redshifts implies that the derived
parameters correspond to different physical scales, for similar
galaxies at different redshift, due to an aperture bias.  This may
induce secondary correlations difficult to address, in particular, if
some of these parameters present a cosmological evolution and/or 
spatial gradients within each galaxy; (iii) while the spectroscopic
information (oxygen abundance and SFR) is derived from this aperture
limited dataset, the third analyzed parameter, the mass, is derived
mostly based on integrated photometric data. Moreover, in some cases
it is derived using the information comprised in the spectroscopic
data to estimate the corresponding M/L-ratio
\citep{mann10,kauffmann03}. Contrary to what is often claimed, we recall that these systematic effects cannot be reduced by the large statistical
number of data encompassed in the considered dataset. 

\begin{figure*}
\centering
\includegraphics[width=8cm,clip,trim=0 20 60 0]{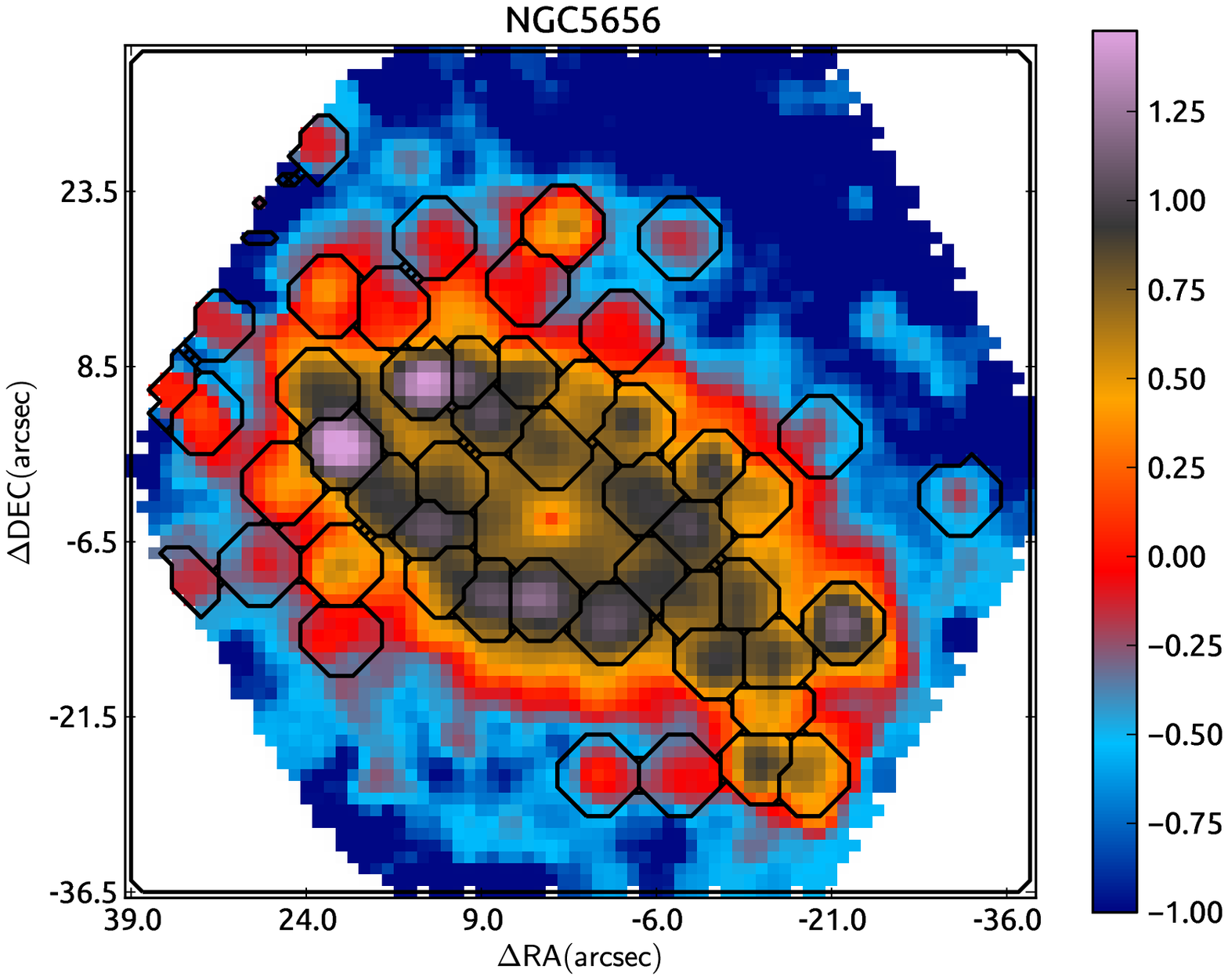}
\includegraphics[width=8cm,clip,trim=0 20 60 0]{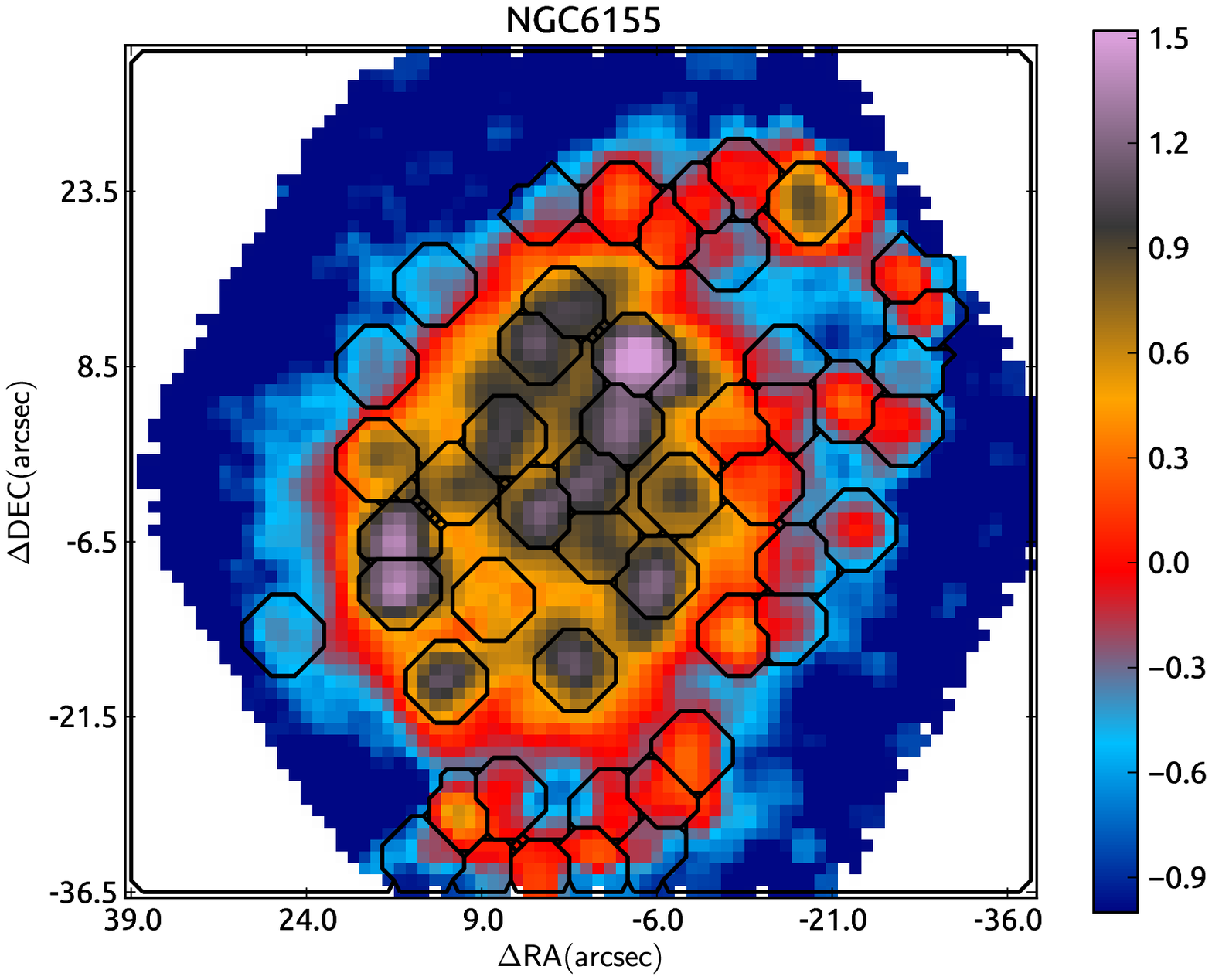}
\includegraphics[width=6.1cm,angle=270,clip,trim=0 0 0 0]{figs/98NGC5656.ps}
\includegraphics[width=6.1cm,angle=270,clip,trim=0 0 0 0]{figs/98NGC6155.ps}
\caption{\label{fig:HII} {\it Top panels:} IFS-based H$\alpha$ maps, in units of \FunitsA, derived for two representative galaxies of the  sample (color images), together with the detected \ion{H}{ii} regions shown as  black segmented contours. {\it Bottom panels:} Radial distribution  of the oxygen abundance derived for the individual \ion{H}{ii}  regions with abundace errors below 0.15 dex, as a function of the deprojected distance (i.e., corrected for inclination), normalized to  the effective radius, for the same galaxies. The size of the circles is proportional to the H$\alpha$ intensity.}
\end{figure*}

Some of these issues are partially solved when a large aperture is
used to derive the spectra of each galaxy, as in the case of
drift-scan observations \citep[e.g.][]{mous10,hugh12}. In this case,
the derived abundance (basically luminosity weighted across the
optical extent of the galaxies) has a better correspondence with
the characteristic one \cite[i.e., the value at 0.4$\rho_{25}$, where
  $\rho_{25}$ is the radius at a surface brightness of 25
  mag/arcsec$^2$, ][]{zaritsky94,Garnett:2002p339}, as demonstrated by
\cite{mous06}.  On the other hand, the SFR can be directly derived from
the integrated H$\alpha$ emission across the considered
aperture. However, these observations have also some limitations: (i)
in many cases they do not cover the entire optical extent of the
galaxies, and the covered fraction is different from galaxy to galaxy,
which introduces again an aperture uncertainty \citep[Fig. 3 of ][
  for an example]{sanchez11}; (ii) due to the lack of spatially
resolved spectroscopic information, they mix regions with different
ionization properties and/or ionization sources, that could change
substantially at different locations even within quiescent spiral
galaxies \citep[e.g.][]{sanchez12b}. This introduces uncertainties in
both the derived abundances and the estimated SFR.

\begin{figure*}
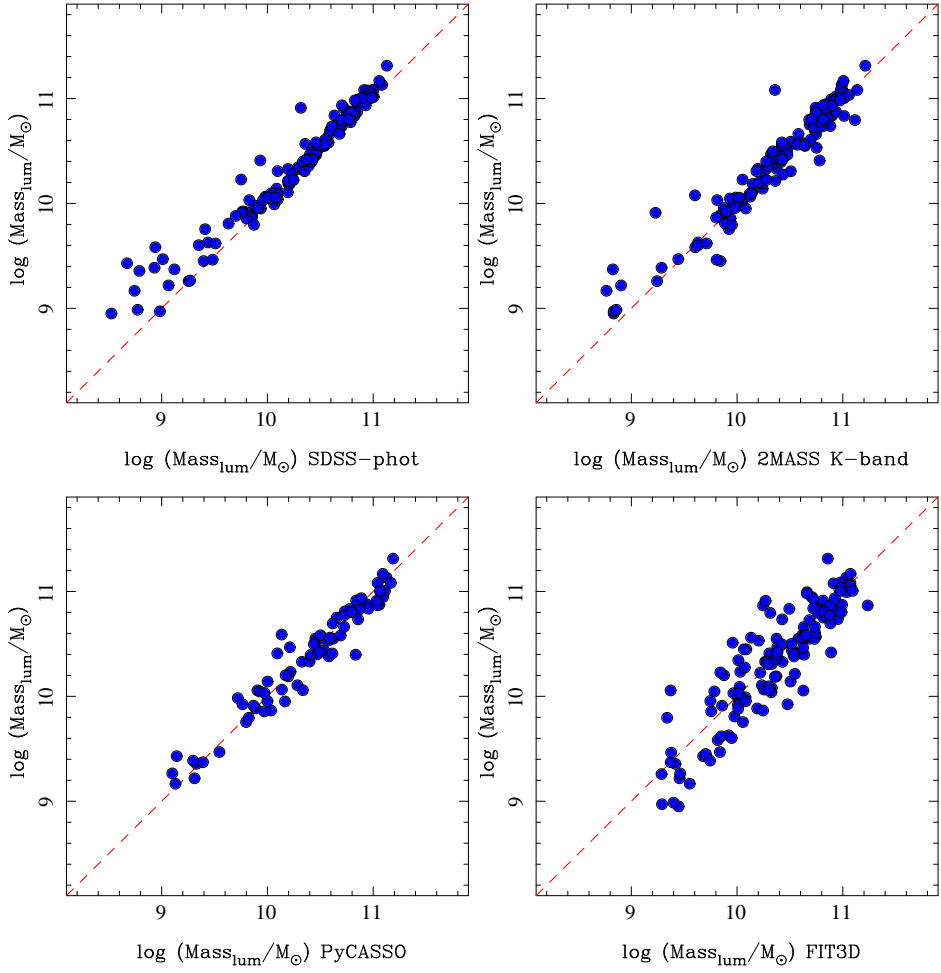

\centering
\includegraphics[width=6.5cm,angle=270,clip,trim=0 0 -20 0]{figs/MASS_SDSS_phot.ps}
\includegraphics[width=6.5cm,angle=270,clip,trim=0 0 -20 0]{figs/MASS_2MASS.ps}
\includegraphics[width=6.5cm,angle=270,clip,trim=0 0 -20 0]{figs/MASS_PyCASSO.ps}
\includegraphics[width=6.5cm,angle=270,clip,trim=0 0 -20 0]{figs/MASS_FIT3D.ps}
\caption{\label{fig:Mass}  Distribution of the stellar masses derived using the average M/L ratios described in \cite{bell01} and the $B$ and $V$-band photometry (Y axis in all the panels) along the stellar masses derived using other procedures (X axis in all the panels): {\it Top-Left panels} Stellar masses derived using all the five-bands SDSS photometry and the M/L ratio provided by the SED analysis performed using PARADISE;  {\it Top-Right panel:} Stellar masses derived using the 2MASS K-band photometry and the M/L ratio provided by \cite{long09}; {\it Bottom-left panel:} Stellar masses derived using the CALIFA datacubes spectrophotometry and the M/L ratio provided by PyCASSO/Starlight; and {\it Bottom-right panel:} Stellar Masses derived from CALIFA datacubes spectrophotometry and the M/L ratio provided by FIT3D}
\end{figure*}

In spite of these limitations, \cite{hugh12} explored in a recent
study the dependence of the \mz\ relation on the SFR, using drift-scan
observations of 135 nearby late-type galaxies in the Local
Universe. As expected, their dispersion around the \mz\ relation is
similar or even slightly higher than the one reported by T04. {
  Although, they cannot reproduce the results by \cite{mann10} and
  \cite{lara10a}, and in fact the scatter increases when they try to
  introduce a dependence on the SFR, they found a strong depedence of
  the metallicity with the gas fraction. This dependece could induce a
  correlation with the SFR in a large sample of galaxies.}

{  More
  recently, \cite{PeMo12} and \cite{cres12} studied the \mz\ relation
  in a wide range of redshifts both using the zCOSMOS data, and the
  SDSS ones in the first case. They found either a negative trend of
  the metallicity with the SFR for a fixed galaxy mass, in the first
  case, or a distribution of masses, metallicies and SFRs consisten
  with the FMR, in the latter one, both in agreement with
  \cite{mann10} and \cite{lara10a}. Despite of this agreement,
  \cite{PeMo12} noted that the dispersion around the \mz\ relation
       {\it is only reduced by 0.01 dex at all stellar masses} when a
       secondary relation with the SFR is introduced, for an initial
       dispersion of $\sim$0.1 dex. As a comparison we should note that
       the decrease of the dispersion reported by \cite{mann10} was
       nearly half of the original value, when introducing the same
       relation. \cite{PeMo12} considered that the lack
  of extreme SFR values in their SDSS galaxies may explain this effect.}


At higher redshifts { the situation is more complicated. \cite{wuyt12}
  found that the distribution of their data is in agreement with the
  local FMR relation presented by \cite{mann10}. However, they do not
  see a correlation between metallicity and SFR at a fixed mass.  On
  the contrary, \cite{rich11} interpret the largest metallicities
  found at the low mass range as a
  consequence of the FMR in combination with a lower star-formation
  rate of these objects. In a similar way, \cite{naka12}
  show that the lower limit to the average metallicities found for
  a sample of Ly$\alpha$ emitters at $z\sim 2.2$, after stacking the
  individual spectra, is roughly consistent with the FMR
  relation. Finally, \cite{magr12}, support the idea that the scaling
  relation between the mass, metallicity and SFR has a
  different origin (and shape) for ``active'' starbursts, more common
  at high redshifts, and quiescent galaxies, more frequent at lower
  ones. The fact that in most cases only the N2 indicator
is accesible at high-redshift may introduce biases difficult to quantify.}

In order to explore the \mz\ relation in the Local Universe, and to
bring light to its dependence on the SFR, while minimizing the bias
effects previously described, we have used the integral-field
spectroscopic (IFS) data provided by the CALIFA survey
\citep{sanchez12a}\footnote{http://califa.caha.es/}. CALIFA is an
on-going exploration of the spatially resolved spectroscopic
properties of galaxies in the Local Universe ($z<$0.03) using
wide-field IFS to cover the full optical extent (up to $\sim$2.5
r$_e$) of $\sim$600 galaxies of any morphological type, distributed
across the entire Color-Magnitude diagram (Walcher et al., in prep.),
and sampling the wavelength range
3650-7500\AA. So far, the survey has completed $\sim$1/3 of its
observations, and the first data release, comprising 100 galaxies {
has been delivered recently \citep{huse13}}.

The layout of this article is as follows: in Sec.
\ref{sample} we summarize the main properties of the sample and data
used in this study; in Sec. \ref{ana}, we present the main
analysis, and the derivation of the parameters analyzed
in the article: the mass, metallicity and SFR for each individual
galaxy and \HII\ region; the global  \mz\ relation and its dependence
with the SFR is explored in Sec. \ref{mz} and \ref{mz_sfr}; a
similar analysis for the local \mz\ (or \Sz) relation is presented in
Sec. \ref{local}; we explore the possible dependence on the specific
star-formation rate (sSFR) in Sec. \ref{sSFR}; our results and
conclusions are summarized in Sec. \ref{summary}.

\section{Sample of galaxies and data}\label{sample}

The galaxies under study have been selected from the CALIFA observed
sample$^1$. Since CALIFA is an ongoing
survey, whose observations are scheduled on a monthly basis (i.e., dark
nights), the list of objects increases regularly. The current
results are based on the 150 galaxies observed using the
low-resolution setup until July 2012. At that point, most of the
color-magnitude diagram had been sampled by the survey with at least
one or two targets per bin of magnitude and color, including galaxies
of any morphological type. The CALIFA mother sample becomes incomplete
below M$_r>-$19 mag, which corresponds to a stellar mass of $\sim$10$^{9.5}$ M$_\odot$.
Therefore, it does not sample low-mass and/or dwarf galaxies.
Most of these galaxies are part of the 1st
CALIFA Data Release \cite{huse13}, and therefore the
datacube are accesible from the DR1 webpage\footnote{\url{http://califa.caha.es/DR1}}.  Table
\ref{table_Mass} shows the list of the galaxies analyzed in the current
study, including, for each galaxy: (i) its name, (ii) redshift, (iii)
$V$-band magnitudes and $B-V$ color and visual morphological
classification (extracted from Walcher et al. in prep).  In addition, we
include the main properties derived along this article, as explained
later: (i) the integrated stellar mass, (ii) characteristic oxygen
abundance and (iii) the integrated star-formation.

The details of the survey, sample, observational strategy and
reduction are explained in \cite{sanchez12a}. All galaxies were
observed using PMAS \citep{roth05} in the PPAK configuration
\citep{kelz06}, covering a hexagonal field-of-view of
74$\arcsec$$\times$64$\arcsec$, sufficient to cover the full optical
extent of the galaxies up to 2-3 effective radii. This is possible due
to the diameter selection of the sample (Walcher et al., in prep.).
The observing strategy guarantees a complete coverage of the FoV, with
a final spatial resolution of FWHM$\sim$3$\arcsec$, i.e., $\sim$1 kpc
at the average redshift of the survey. The sampled wavelength range
and spectroscopic resolution (3745-7500\AA,
$\lambda/\Delta\lambda\sim$850, for the low-resolution setup) are more
than sufficient to explore the most prominent ionized gas emission
lines, from [OII]$\lambda$3727 to [SII]$\lambda$6731, on one hand, and
to deblend and subtract the underlying stellar population, on the
other hand \citep[e.g.][Cid Fernandes et al.,
  submitted]{sanchez12a,kehrig12}. The dataset was reduced using
version 1.3c of the CALIFA pipeline, whose modifications with respect
to the one presented in \cite{sanchez12a} are described in detail in
\cite{huse13}. In summary, the data fulfill the foreseen quality
control requirements, with a spectrophotometric accuracy better than a
15\%\ everywhere within the wavelength range, both absolute
and relative, and a depth that allows us to detect emission lines in
individual \HII\ regions as weak as $\sim$10$^{-17}$\fluxA, with a
S/N$\sim$3-5.

\begin{figure}
\centering
\includegraphics[width=6.5cm,angle=270,clip,trim=0 0 -20 0]{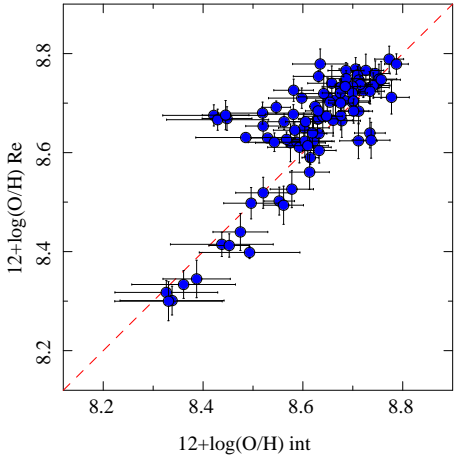}
\caption{\label{fig:int_Re}  Distribution of the oxygen abundances estimated for each galaxy at the effective radius (Re), using the radial gradients, and the oxygen abundances derived directly from the spectra integrated over the entire field-of-view (int).}
\end{figure}

The final product of the data-reduction is a regular-grid datacube,
with $x$ and $y$ coordinates indicating the right-ascension and
declination of the target and $z$ being a common step in
wavelength. The CALIFA pipeline also provides the propagated
error cube, a proper mask cube of bad pixels, and a prescription of
how to handle the errors when doing spatial binning (due to covariance
between adjacent pixels after image reconstruction). These datacubes,
together with the ancillary data described in Walcher et al. (in
preparation), are the basic starting point of our analysis.

{ In addition to these galaxies, we have included { 31 out of
    the 38} face-on spirals analyzed in \cite{sanchez12b}. This sample
  comprises face-on spiral galaxies extracted from the CALIFA
  feasibility studies \citep{marmol-queralto11}. All these galaxies
  were observed with PPAK, using a setup similar to the one used for
  CALIFA, with slightly different spectral resolutions and wavelength
  coverage \citep[see][ for details]{marmol-queralto11}. They cover
  the same redshift range, optical size and the extent covered by the
  CALIFA, too.  On the other hand, the remaining 7 spirals were
  extracted from the PINGS survey \citep{rosales-ortega10}. These
  galaxies are located at lower redshift, being larger in projected
  sizes, and in some cases they have not been completely covered by
  the IFS data.  Therefore, we prefer to excluded them from this
  analysis.}


\section{Analysis and Results}\label{ana}

\subsection{Derivation of the analyzed parameters}\label{ana_par}

The three main quantities to derive for each galaxy are (i) the
integrated stellar mass, (ii) the characteristic oxygen abundance
and (iii) the integrated SFR.  These parameters were derived following
the procedure described in \cite{sanchez12b}, developed for a similar
set of data. In the following we briefly outline
the most important aspects of their derivation.

\subsubsection{Stellar Masses}\label{mass}

We derive the stellar masses using the
integrated $V$-band magnitudes and $B-V$ colors listed in Table
\ref{table_Mass}, and the average mass-luminosity ratio ($M/L$)
described in \cite{bell01}. { The $B$ and $V$-band magnitudes were
derived using the Petrosian magnitudes provided the SDSS photometric
catalog \footnote{http://www.sdss.org/dr7/access/}, and transformed
using the equations by \cite{jest05}. The errors provided by the
SDSS catalog were propagated to derive the corresponding errors for
the $B$ and $V$-band magnitudes. The expected systematic error of
the transformation, of $\sim$0.02 mag, was not taken into account
since it will affect equally to all the listed values. }.
{ A direct comparison with our own derived magnitudes, based on a growth-curve
analysis and a detail subtraction of the local background of the SDSS images
shows that the errors could be {\it ad maximum} $\sim$0.07 mag (Walcher et al., in prep.)}
 The integrated luminosities and colors
were corrected for the effect of dust attenuation, where the internal
extinction was derived from the multi-SSP analysis of the stellar
continuum summarized in \cite{sanchez12a} { , prior to derive the stellar masses}.

{ This procedure to derive the stellar masses is robust and straight-forward,
and it only requires a few assumptions on the properties of the galaxies. 
We adopt the \cite{bell01} $M/L$ ratios, instead of the more recent ones provided by \cite{bell03} 
for the SDSS colors, for consistency with the derivation performed in \cite{sanchez12b}.
Indeed, \cite{bell03} show that both $M/L$ ratios are consistent with each other. 

In any case, we perform a sanity check to test the
robustness of the stellar masses derived. For doing so, we derive the
masses using four additional methods: (i) we use the full spectral
energy distrubution (SED) provided by the SDSS photometry, and we
derive the masses using the stellar population analysis provided by
the PARADISE code \citep{walcher11}. This derivation is less affected
by possible color biases, since it includes the whole optical
SEDs. However, the dust attenuation affecting the stellar populations
was not taken into account; (ii) we derive the stellar masses using
the K-band photometry extracted from the 2MASS photometric
catalog \footnote{http://www.ipac.caltech.edu/2mass/releases/allsky/},
applying a global aperture correction, and adopting the M/L ratio
described by \cite{long09}. This derivation is clearly the one that is
less affected by any possible color biases, the effect of young
stellar populations and dust effects, a priori. However, it presents its
own problems, in particlar the uncertainties in the M/L ratio related
to the unknown contribution of the TP-AGB stars in this wavelength regime \citep[e.g.][]{walcher11}. Only 128 objects from
the ones considered here have photometry in the 2MASS catalog; (iii)
we collected the stellar masses recently published by \cite{perez13},
extracted from the CALIFA datacubes, using {\sc p}y{\sc casso} (Cid
Fernandes et al., submitted), the 3D implementation of the Starlight
code \citep{cid-fernandes05}, for those galaxies in common (75
objects). These masses are the most self-consistent with the other
parameters presented in this article, but they rely on a less
straight-forward stellar population decomposition. The analysis takes
into account the dust attenuation; and (iv) we derived the stellar
masses provided by the stellar population decomposition method used to
remove the underlying stellar population pixel-by-pixel, using FIT3D.
This method is, in principle, similar to the previous one. However,
FIT3D uses a much more reduced library of stellar populations, and
therefore, it provides, a priory, a less accurate M/L ratios and
stellar masses.

Figure \ref{fig:Mass} illustrates the results of this analysis. For each
panel, it shows the comparison between the stellar mass derived using
the \cite{bell01} relation, and those ones described before. In all
the cases we find a very good agreement, close to a one-to-one relation
in most cases. The best agreement is with the stellar masses
derived using only photometric information, either the full SDSS SED
distribution or the 2MASS K-band data, with a standard deviation of
the difference of $\sigma\sim$0.15 dex. The masses derived using only
the SDSS photometry are slightly lower, in particular in the low mass
regime, which is expected since they are not corrected by dust
attenuation. For the stellar masses derived using the CALIFA
spectroscopic information, the best agreement is with those provided by {\sc p}y{\sc casso}. The standard deviation of the difference is $\sigma\sim$0.14
dex, which is consistent with an spectrophotometric accuracy of the CALIFA datacubes
of $\sim$10\% with respect to the SDSS one. Finally, the largest dispersion
is found when comparing the adopted stellar masses with those derived
from the multi-SSP analysis provided by FIT3D ($\sigma\sim$0.24 dex), as expected.}

Similar results are found when comparing the masses derived with the
total stellar masses listed in the MPA/JHU catalog \cite{kauff03a}.
Therefore, if there is a systematic effect in the derivation of the
stellar masses, this is { at maximum of the order of} the typical error estimated based
on the propagation of the photometric errors ($\sim$0.15 dex).

\begin{figure*}
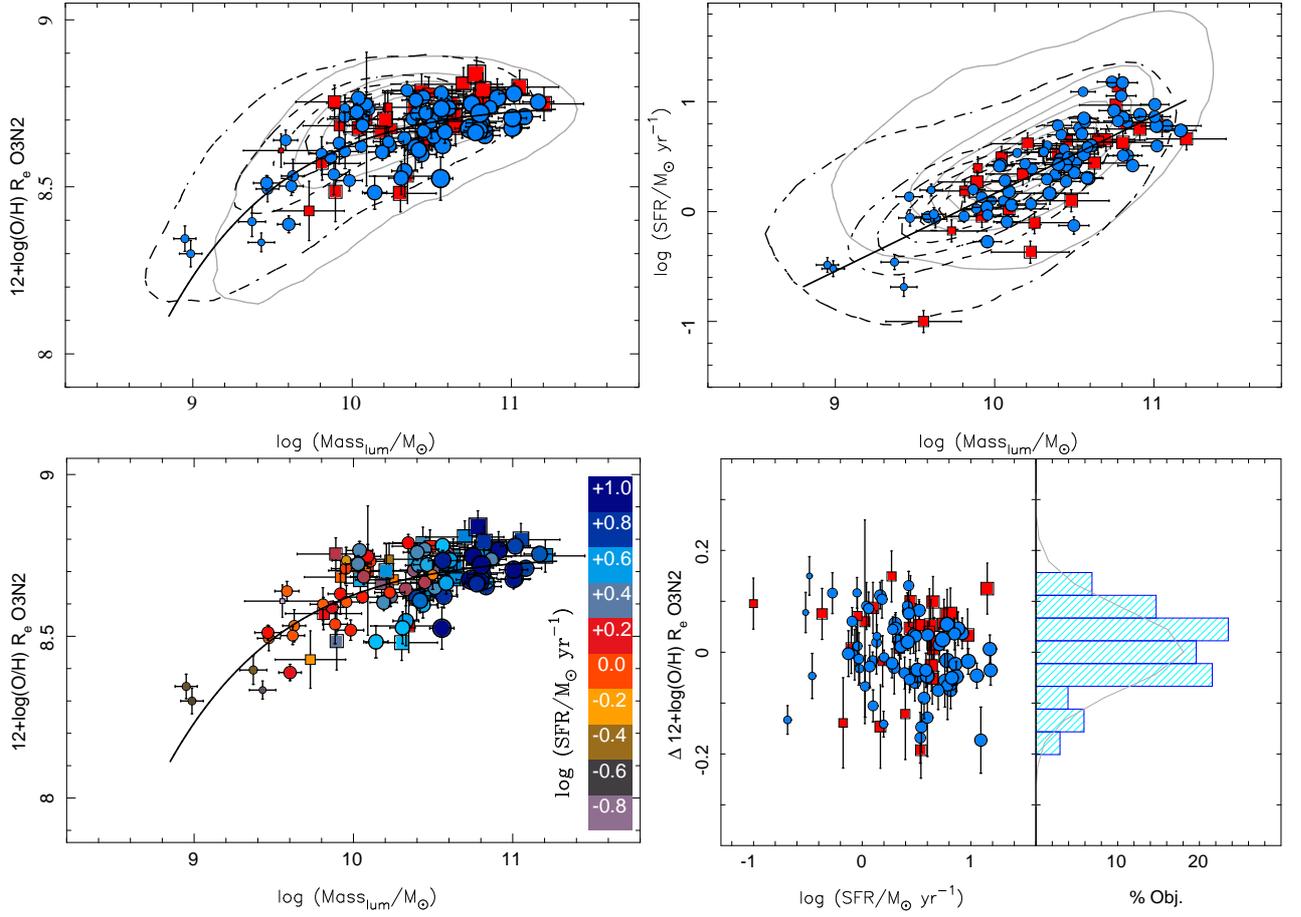

\centering
\includegraphics[width=6cm,angle=270]{figs/7Mass_OH_9.ps}
\includegraphics[width=6cm,angle=270]{figs/9Mass_OH_9.ps}
\includegraphics[width=6cm,angle=270]{figs/b7Mass_OH_9.ps}
\includegraphics[width=6.2cm,height=8.2cm,clip,trim=0 0 0 0,angle=270]{figs/18Mass_OH_9.ps}
\caption{\label{fig:MZ}. {\it Top-left panel:} Distribution of the oxygen abundances at the effective radii as a function of the integrated stellar masses for the CALIFA galaxies (113, blue solid circles). For comparison purposes, we show similar values for the galaxies observed in the CALIFA feasibility studies (31, red solid squares) included in the \HII regions catalog described in \cite{sanchez12b}. The size of the symbols is proportional to the integrated SFR. The solid line represents the best fitted curve, as described in the text. We also included the distribution of abundaces as a function of the stellar masses for the SDSS data corresponding to the redshift ranges sampled by \cite{lara10a}, dashed black contours, and \cite{mann10}, solid grey contours. In both cases, the first contour encircles the 95\% of the objects, and 20\% less objects in each consecutive contour. {\it Top-right panel:} Distribution of the integrated SFR as a function of the stellar masses. The size of the symbols is proportional to the oxygen abundance shown in previous panel. The solid line represents the best fitted linear regression to the CALIFA data. { Like in the previos panel the contours represent the distribution of both parameters corresponding to the redshift ranges sampled by \cite{lara10a}, dashed black contours, and \cite{mann10}, solid grey contours }{\it Bottom-left panel:} Same distribution shown in the top-left panel, for the CALIFA and feasibility-studies objects. The colors represent the logarithm of the integrated SFR for each galaxy. {\it Bottom-right Panel:} Distribution of the differential oxygen abundances at the effective radii, once the dependency with the stellar mass has been subtracted, as a function of the integrated SFR for the CALIFA galaxies (113, blue solid circles). The size of the symbols is proportional to the oxygen abundances shown in Fig. \ref{fig:MZ}. The histogram shows the same distribution of differential oxygen abundances. The solid line represents a Gaussian function with the same central value (0.01 dex) and standard deviation (0.07 dex) as the represented distribution, scaled to match the histogram. }
\end{figure*}

\subsubsection{Star Formation Rate}\label{SFR}

The SFR was derived for each galaxy based on the properties of the
ionized gas emission. By construction, CALIFA IFS data sample most of
the optical extent of galaxies, which allows to derive the total SFR
minimizing any aperture effect. The SFR was derived from the
integrated H$\alpha$ luminosity.  For doing so, we first perform a
spectroscopic decomposition analysis between the underlying stellar
population and the ionized gas emission lines described using FIT3D
\citep{sanchez07b,rosales11,sanchez12a,sanchez12b}. The analysis was
performed spaxel-by-spaxel, following \cite{sanchez11}.  Each spectrum
in the datacube is then decontaminated by the underlying stellar
continuum using the multi-SSP model derived .  Then, we selected those
spaxels whose ionization is dominated by star-formation based on
emission line ratios \cite[e.g.][]{sanchez12b}, i.e., the
\HII\ regions described latter. The areas dominated by diffuse gas
emission have been excluded on purpose, since their nature is still
unclear { \cite[e.g.][, Iglesias-P\'aramo et al., in prep.]{thilk02}}. This does not affect
the results, since diffuse gas comprises less than 5\% of the total
integrated flux of H$\alpha$ for our galaxies. We repeated all the 
calculations using the SFR integrating over all the FoV, instead of just
only the \HII\ regions without any significant difference. The observed H$\alpha$
intensities were corrected for reddening using the Balmer decrement
($H\alpha/H\beta$) according to the reddening function of
\citet{cardelli89}, assuming $R \equiv A_V/E(B-V) = 3.1$. The theoretical
value for the intrinsic Balmer line ratios were taken from
\citet{osterbrock89}, assuming case B recombination (optically thick
in all the Lyman lines), an electron density of $n_e = 100$ cm$^{-3}$
and an electron temperature $T_e = 10^4$ K, which are reasonable
assumptions for typical star-forming regions. The values of the SFR
were derived for each galaxy adopting the classical relations by
\citet{kennicutt98}, based on the dust-corrected H$\alpha$
luminosities.

{ The corresponding errors for the SFR were derived, for each
  galaxy, by propagating through the equations the errors provided by
  the fitting procedure for the derivation of H$\alpha$ and H$\beta$
  emission lines, spaxel-to-spaxel
  \cite[FIT3D,][]{sanchez11,sanchez12b}.  These errors do not include
  the possible zero-point calibration errors in the CALIFA datacubes,
  which are estimated to be lower than a $\sim$15\% through all the
  wavelength range \citep{huse13}.
}

\subsubsection{Oxygen Abundance}\label{OH}

The characteristic oxygen abundance was defined by \cite{zaritsky94}
as the abundance at 0.4$\rho_{25}$, being representative of the
average value across the galaxy, as we indicated before. This distance
corresponds basically to one effective radius for galaxies in the local
universe. We demonstrated in \cite{sanchez12b}, that the effective
radius is a convenient parameter to normalize the abundance gradients
in galaxies at different redshifts. We will address the study of
abundance gradients and the relation between the integrated and characteristic 
oxygen abundance in a forthcoming article (S\'anchez et al. in prep.). 
Thus, we re-define the characteristic or effective oxygen abundance as
the corresponding value at the effective radius.

To derive this parameter we analyze the radial gradients of the oxygen
abundance in all the galaxies of our sample, using the procedures
detailed in \cite{sanchez12b}. In a companion article we will describe
the procedure for this particular dataset, summarizing the main
properties of the ionized regions across the galaxies (Sanchez, in
prep.).  We present here just a brief summary of the different steps
included in the overall process: (i) First we create a narrow-band
image of 120$\AA$ width, centered at the wavelength of H$\alpha$
shifted at the redshift of the targets. The narrow-band image is
properly corrected for the contamination of the adjacent continuum;
(ii) The narrow band image is used as an input for
\textsc{HIIexplorer}\footnote{\url{http://www.caha.es/sanchez/HII_explorer/}},
an automatic \ion{H}{ii} region detection code created for
this kind of analysis \citep{sanchez12b,rosales12}. The code provides
with a segmentation map that identifies each detected ionized
region. Then, it extracts the integrated spectra corresponding to each
segmented region. Figure \ref{fig:HII} illustrates the process,
showing the H$\alpha$ intensity maps and the corresponding
segmentations, for two objects. A total of 3435 individual ionized
regions are detected in a total of 137 galaxies from the sample; (iii)
Each extracted spectrum is then decontaminated by the underlying
stellar continuum using the multi-SSP fitting routines included in
FIT3D, as described before; (iv) Then each emission line within the
considered wavelength range is fitted with a Gaussian function to
recover the line
intensities and ratios; (v) These line ratios are used to discriminate
between different ionization conditions and to derive the oxygen
abundances for each particular ionized region. { Basically we selected
those clumpy ionized regions with a strong blue underlying stellar population,
that contributes at least a 20\% of the flux in the $V$-band,
to ensure that they are located below the \cite{kewley08} demarcation line
in the classical [OIII]/H$\beta$ vs. [NII]/H$\alpha$ diagnostic diagram \citep{baldwin81}}.
A total of 2846
\ion{H}{ii} regions are then extracted, for a total of 134 galaxies;
(vi) Finally, in combination with a morphological analysis of the
galaxy, we can recover the abundance gradients for each particular
galaxy, and derive the corresponding value at the effective radius.
{ For doing so, we fit the radial distribution of the oxygen abundance between
0.2 and 2.1 effective radius with a linear regression, and uses the derived zero-point and 
slope to determine the oxygen abundance at one effective radius. This procedure
increses the accuracy of the derived abundance, since it uses the full radial distribution
of individual estimations.} We
used only (1) those regions for which the oxygen abundance was derived
with a nominal error lower than $<$0.2 dex, which corresponds to
emission lines with a S/N$>$5 in all the considered emission lines
{ ([OII]3727, H$\beta$, [OIII]5007, H$\alpha$ and [NII]6583)}, and (2) those
galaxies with at least 3 \HII\ regions covering at least between 0.3
and 2.1 effective radii. This comprises a total of 2061 \ion{H}{ii}
regions and associations, distributed in 113 galaxies. This sample
comprises galaxies of any type, mostly spirals (both early and late
type), with and without bars, and with different inclinations. We note
that although the detectability of \HII\ regions is affected by
inclination, i.e., less number is accesible and detected, we do not
notice any further difference in their properties (abundance, radial
distribution, luminosity...).

The oxygen abundance was derived using different strong-line
indicators, like the R23-fit derived by T04, the N2 and O3N2
calibrators by \cite{pettini04}, and the recently {\it
  counterpart}-method by \cite{P12}. Interestingly, despite the differences
in the derived estimates, all values show a 
trend with the stellar mass \citep[e.g.][]{kewley08}. In fact, there is a
clear correspondence not only with the calibrators
\citep[e.g.][]{angel12}, but also between the line ratios used to
derive them \citep[e.g.][]{sanchez12b}. Therefore, for simplicity, we
will show here only the results based on the O3N2-calibrator, which
is one of the most straight-forward one.

The results of this analysis are listed in Table \ref{table_Mass},
including the final number of \ion{H}{ii} regions considered for each
galaxy and the derived parameters, i.e., the integrated stellar mass
and SFR, and characteristic oxygen abundance { , together with their corresponding errors}. In addition to these
characteristic values for each individual galaxy, we derive similar
parameters for each individual \ion{H}{ii} region: the mass surface
density ($\Sigma$) and the surface SFR density ($\mu$SFR). These quantities were
derived using the same procedures described in Sec. \ref{mass} and
\ref{SFR}, but using as observables (i.e., the photometric values and
the H$\alpha$ fluxes), those ones corresponding to each particular
\HII\ region.  Finally, the derived values were divided by the
corresponding encircled area.



\begin{figure*}
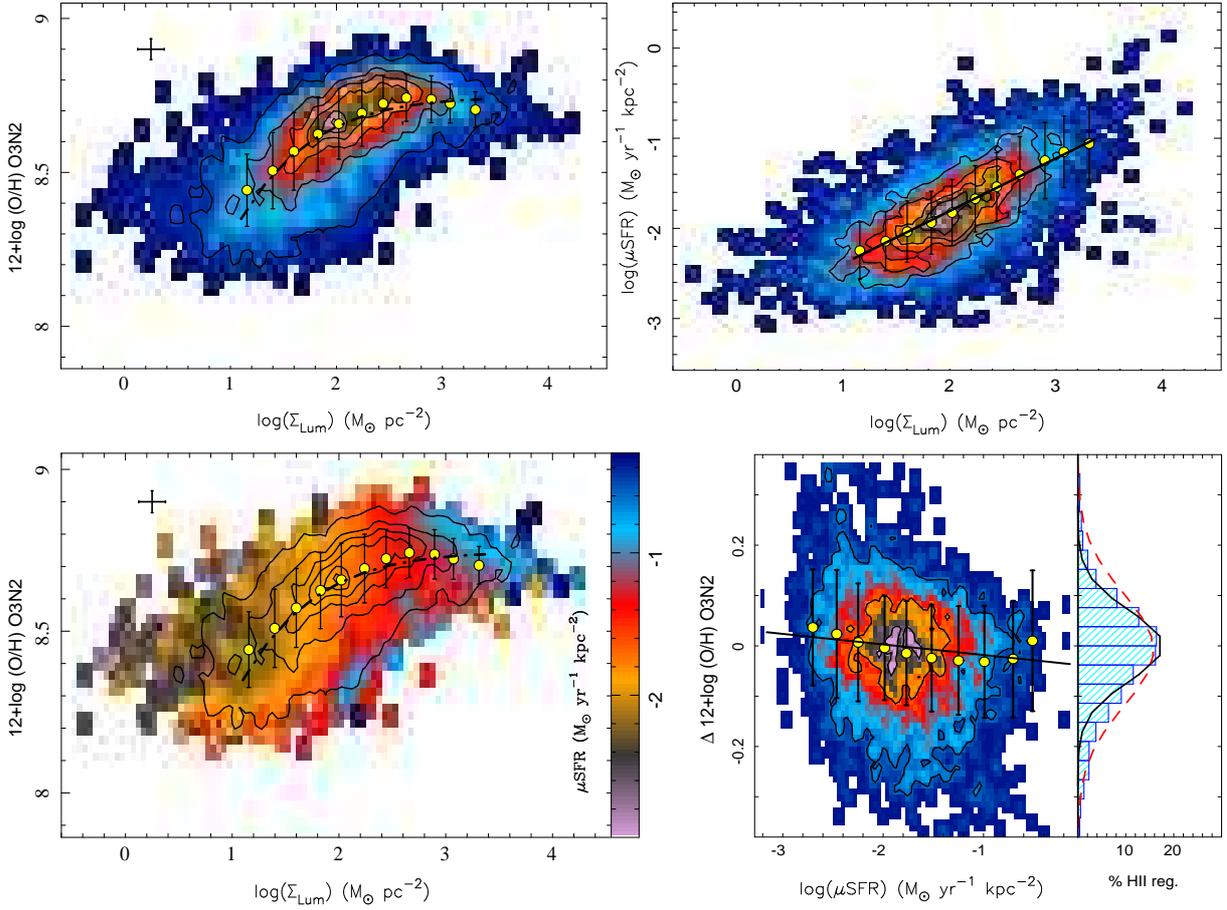

\centering
 \includegraphics[height=8cm,clip,trim=0 0 0 0,angle=270]{figs/OH_lM_0.ps}
 \includegraphics[height=8cm,angle=270]{figs/local_lM_SFR.ps}\vspace{0.2cm}
 \includegraphics[width=6cm,height=9cm,angle=270]{figs/local_lM_OH_SFR.ps}
 \includegraphics[height=7cm,width=6cm,clip,trim=0 0 0 0,angle=270]{figs/SFR_Delta_OH.ps}
\caption{\label{fig:MZ_local}. {\it Top-left panel:} Distribution of the oxygen abundances  for the 3000 individual \HII regions extracted from the CALIFA data, as a function of the surface mass density, represented with a density contour-plot. The first contour encircles 95\% of the total number of \HII regions, with $\sim$20\% less in each consecutive contour. The solid-yellow points represent the average oxygen abundances, with their corresponding standard deviations indicated as error bars, for consecutive bins of 0.1 dex in surface mass density. The dot-dashed line represents the best fitted curve for these points. {\it Top-right panel:} Distribution of the SFR surface density as a function of the stellar mass density of each individual \HII region. The contour plots represent the density of points in the distribution of oxygen abundances as a function of the surface SFR densities, with the same encircled numbers as in the previous panel. {\it Bottom-left panel:} Same distribution shown in the top-left panel, for the CALIFA and feasibility-studies objects. The colors represent the logarithm of the integrated surface density of the SFR for each \HII region. {\it Bottom-right panel:}  Distribution of the differential oxygen abundance, once the dependency with the surface mass density has been subtracted, for individual \HII regions. The contour plots represent the density of points in the distribution of differential oxygen abundances, once subtracted the dependence on the surface masses densities, as a function of the surface SFR densities, with the same encircled numbers as in the previous panel. The solid-yellow points represent the average values for the abundance offsets, with their corresponding standard deviations indicated as error bars, for consecutive bins of 0.1 dex in surface SFR density. }
\end{figure*}

\subsection{Mass Metallicity Relation}\label{mz}

We first explore the \mz\ relation following T04 and
\cite{mann10}. Figure \ref{fig:MZ} shows the distribution of the
characteristic oxygen abundances as a function of the stellar masses
for the galaxies of our sample (blue circles). In addition, we have
included those values corresponding to the galaxies analyzed in
\cite{sanchez12b} from the CALIFA feasibility studies (red squares). As expected, there is a
clear trend between both quantities, following basically the shape
described by \cite{kewley08} for the considered abundance
calibrator. The correlation coefficient between both quantities is
$r$=0.769, indicating that there is a positive trend. From the formal
point of view, this coefficient should be used only for linear
relations. However, it is still valid as an indication of a trend
between two parameters.

For comparison purposes, we include the density distribution of galaxies within the \mz\ plane
for the SDSS galaxies corresponding to the redshift range studied by
\cite{lara10a} (dashed black contours) and \cite{mann10} (solid grey
contours). The first contour encircles
$\sim$95\% of the total number of galaxies, with a 20\% decrease in
each consecutive one. { We have calculated the oxygen abundances
  for these galaxies using the O3N2 indicator, instead of adopting
  published metallicities which will force us to adopt a correction.}
For doing so, the stellar masses and emission line ratios have been
{ directly} extracted from the Max-Planck-Institute for
Astrophysics-John Hopkins University (MPA-JHU) emission-line analysis
database\footnote{http://www.mpa-garching.mpg.de/SDSS}. The contours
show that the CALIFA data cover a similar range of masses and
metallicities similar to both previous studies. However, it is
important to note here that \cite{lara10a} samples a lower mass and
higher metallicity range than \cite{mann10}, which is a clear
effect of the redshift range covered by both studies and the SDSS
selection function.

Instead of the classical polynomial function adopted by
most of previous studies for this relation \citep[e.g.,
  T04,][]{kewley08,mann10,rosales12,hugh12}, we prefer to adopt an
asymptotic function to describe the relation between both parameters
following \cite{mous11}:

{

\begin{equation}\label{eq1}
 {\rm y}=a+b({\rm x}-c)\exp[-({\rm x}-c)]
\end{equation}

}

Where $y =$ 12+log(O/H), and $x$ is the
logarithm of the stellar mass in units of 10$^8$M$_{\odot}$. This functional form tries to
describe the distribution of parameters defining a maximum abundance
for large masses ($a$). Although the adopted formula is slightly
different for computational convenience than the one adopted by
\cite{mous11}, the main motivations and interpretation are the
same. Obviously, due to the sample construction, we do not have enough
data in the low-mass/low-metallicity range to { explore this
second range}. 



The derived parameters for the CALIFA data are $a$=8.74$\pm$0.01,
$b$=0.018$\pm$0.007 and $c$=3.5$\pm$0.3. These two latter parameters
govern how the metallicity actually increases with mass, while the first
one is the asymptotic oxygen abundance for large masses { at the effective radius
of the galaxies}. The derived
values cannot be extrapolated to masses lower than 10$^9$ M$\odot$,
not covered by the CALIFA mother sample. 
A recent
compilation by \cite{kudri12} of the characteristic metallicities on
a set of local universe galaxies indicate that while at high mass they
reach an asymptotic value, at low masses the relation is almost linear,
consistent with early derivations \citep[e.g.][]{skill89,skill92}.
{ With the adopted formula it is possible to reproduce this
  behavior, although the parameters should be adjusted. }

From a theoretical point of view the existence of an asymptotic value
is predicted, as proceeding from the production of elements in stars
and a given IMF, that is, the stellar true yield for a single stellar
population which will be reached when all stars die. { An abundance at
the effective radius of $\sim$8.74 dex corresponds to a maximum abundance
of $\sim$8.84, in the central regions, in agreement with theoretical expectations}. 
For example, \cite{molla05} found a saturation level in 12+log(O/H)=8.80.
They used a grid of chemical evolution models for spiral
and irregular galaxies, the stellar yields by \cite{woos95}
for oxygen and the IMF by \cite{ferr92}. If a different combination of IMF and
stellar yield set is used, differences of a factor two may be
found. An average value of 8.80$\pm$0.15 would be the maximum oxygen
abundance predicted theoretically, which is is also in agreement
with more recent estimations \cite{pily07}.

Since abundance is a relative parameter (amount of oxygen relative to
the amount of hydrogen), this value can also be interpreted as the
maximum abundance of the universe at a certain redshift. As shown by
\cite{mous11}, this parameter evolves with redshift at about $\frac{\Delta log(O/H)}{\Delta z}=-0.2$.  The functional form
describes well the considered data, as it can be appreciated in
Fig. \ref{fig:MZ}. The dispersion of the abundance values along this
curve is $\sigma_{\Delta{\rm log(O/H)}}=$0.07 dex, much smaller than
previous reported values by \citep[e.g., $\sim$0.1 dex,
  T04][]{mann10,hugh12}. It is important to note here that this
dispersion is not much larger than the estimated errors for the
characteristic abundances ($e_{\rm log(O/H)}\sim$0.06 dex). Although
the dispersion around the correlation depends on the actual form
adopted for the \mz\ relation, the derived value is always low. If
instead of the considered functional form we adopt a 3 or 4 order
polynomial function, as the one considered by \cite{kewley08}, the
dispersion is just 0.084 dex, smaller than the value reported in that
publication for SDSS data using the same calibrator. In any case, our
adopted functional form has a more straighforward physical interpretation than
the pure polynomial functions adopted in the literature.

The results do not vary significantly when using the oxygen abundance
derived from the integrated spectra across the covered FoV. However,
both the dispersion around the derived \mz\ relation ({ $\sim$0.1 dex }) and the estimated
errors for the abundances are slightly larger ({ $\sim$0.07 dex}). This is expected since
for the integrated spectra we coadd regions with different ionization
sources \citep[e.g.][]{sanchez12b}, on one hand, and we have a single
estimation of the abundance per object, not a value derived from the
analysis of a well defined gradient (which reduces the error in the
derivation of the oxygen abundance). { Figure \ref{fig:int_Re}
illustrates this effect, showing the distribution of the oxygen
abundances estimated for each galaxy at the effective radius, using
the radial gradients, with the oxygen abundances derived directly
from the spectra integrated over the entire field-of-view. As
expected both parameters shows a tight correlation ($\sigma\sim$0.06
dex). The errors estimated for the integrated abundance are larger,
and they are  directly transfereed to the dispersion along the one-to-one
relation, and thus, to the corresponding \mz\ distribution.}

Once this functional form had ben derived, we determined the offset and
dispersion for the data corresponding to the galaxies analyzed in
\cite{sanchez12b}. Based also on IFS, and covering the full optical
extent of the galaxies, the only possible differences would be
either the details of the  sample selection and/or the redshift range. The
galaxies from the feasibility studies considered here are all
face-on spiral galaxies, with an average redshift similar to that of the
CALIFA data ($z\sim$0.016), and only slightly larger projected sizes
(red squares in Fig. \ref{fig:MZ}). They present a minor offset with
respect to the derived \mz\ relation ($\Delta log(O/H)=-$0.003 dex) and a
slightly larger dispersion ($\sigma_{\Delta log(O/H)}=$0.084
dex). Considering that the spectrophotometric calibration of the data
is not that accurate \citep[$\sim$20\%,][]{marmol-queralto11}, the slightly larger
redshift range covered by this sample \citep{sanchez12b}, and that
the sample is smaller (31 objects), the difference is not
significant. If we consider both this dataset and the CALIFA one described before, the derived dispersion is just $\sigma_{\Delta log(O/H)}=$0.07.


\begin{figure*}
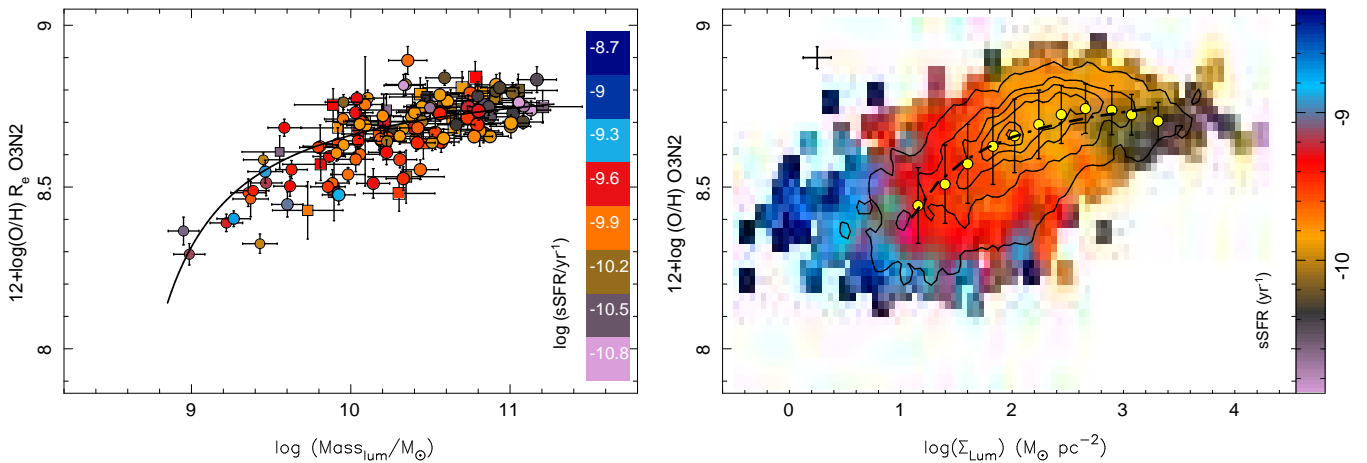

\centering
\includegraphics[width=6cm,angle=270]{figs/c7W_Mass_OH_9.ps}
\includegraphics[width=6cm,angle=270]{figs/local_lM_OH_sSFR.ps}
\caption{\label{fig:MZ_sSFR} {\it Left panel:} Same distribution of
  the characteristic oxygen abundance as a function of the stellar mass as shown in
  the bottom panel of Fig. \ref{fig:MZ}. {\it Right
    panel:} Same distribution of the oxygen abundances
  as a function of the stellar surface density for the individual \HII\ regions,
  as shown in the bottom panel of Fig. \ref{fig:MZ_local}. In both cases, the color
  code indicates the logarithm of the specific star formation rate (sSFR).
}
\end{figure*}

\subsection{Dependence of the \mz\ relation on the SFR}\label{mz_sfr}

The main goal of this study is to explore whether there is a dependence of
the \mz\ relation on the SFR. Before doing that, it is important to
understand the possible additional dependence of the SFR on other
properties of the galaxies, in particular on the stellar mass.  The
SFR of blue/star-forming galaxies has a well known relation with the
stellar mass which has been described not only in the Local Universe
\citep[e.g.][]{brin04}, but also at different redshifts \citep[e.g.][
  and references there in]{elbaz07}. Its nature is still under debate
\citep[e.g.][]{dave08,vulca10}, but it is most probably related to the
ability of a certain galaxy to acquire neutral gas, together with the
well known Schmidt-Kennicutt law \citep{kennicutt98}. Regardless of its nature,
this relation, together with the \mz\ relation may induce a
secondary correlation between the SFR and the oxygen abundance.

Figure \ref{fig:MZ}, top-right panel, shows the distribution of the
SFR as a function of the stellar mass for both the CALIFA galaxies
analyzed here and the galaxies extracted from the feasibility
studies. As expected, both parameters show a clear correlation, with a
strong correlation coefficient ($r=$0.77). The slope of this relation,
for our galaxies, is $\alpha=$0.63, only slightly smaller than the
value reported for the Local Universe based on SDSS data
\citep[$\sim$0.77,][]{elbaz07}. This value is somewhat smaller than
the value near to one expected from the simulations
\citep{elbaz07}. The limited number of low-SFR objects in our sample,
may explain the difference in the slope { . Nevertheless it is clear that we reproduce the expected correlation }. { In order to investigate if the
  selection criteria of the CALIFA galaxies, or the limited number of
  observed galaxies, have induced any spurious correlation that may
  affect the distribution of both quantities and/or the described
  trend, we have compared these distribution with those of the
  complete SDSS sample for the redshift ranges covered by
  \cite{lara10a} and \cite{mann10}. For doing so, we uses the MPA-JHU
  catalog described in the previous section, extracting the same
  stellar masses and the aperture corrected SFR for both redshift
  ranges. The distribution of both parameters were represented in
  Fig. \ref{fig:MZ}, top-right panel, using similar density contours
  as the ones shown in the top-left panel. It is clear that the ranges
  of both quantities, and the distribution along the Mass-SFR plane,
  agree between the three samples. In particular, the agreement is
  remarkable between our analyzed CALIFA sample and the one studied by
  \cite{lara10a}, which is the one at more similar redshift
  range. Indeed, the described correlation agrees completely with the
  SDSS data corresponding to these redshift range.  Thus, the
  selection criteria does not seem to affect neither the range of
  parameters nor the correlation between both quantities. }

Figure \ref{fig:MZ}, bottom-left panel, shows the \mz\ relation, 
but with a different color-coding
indicating the integrated SFR of each
particular galaxy. There is a trend of the SFR with the
metallicity, which follows the trend of the SFR with
mass. Thus, more massive galaxies have both higher SFR and are more
metal rich. In order to determine which is the dominant correlation,
we derive the correlation coefficient between the SFR and the oxygen
abundance, obtaining a value of $r=0.54$. This indicates that, while
the strength of the \mz\ and the SFR-Mass relations are
indistinguishable statistically, the SFR-abundance one is
weaker, and seems to be induced by the other two.

We do not find any trend between the SFR and the oxygen
abundance for a given mass, as the one described by \citep{lara10a}
and \cite{mann10}. For a given mass, the galaxies with stronger
star-formation do not seem to have a lower metal content. In order to
investigate it further, we derived the differential oxygen
abundance with respect to the considered \mz\ relation, to search for 
any correlation with the SFR.

Figure \ref{fig:MZ}, bottom-right panel, shows the distribution of the
residual of the oxygen abundance, once the \mz\ relation has been
subtracted, as a function of the SFR. The data is consistent with being randomly
distributed around the zero value, with no correlation
between both parameters ($r=0.08$). The histogram of the differential
oxygen abundances shows a distribution compatible with a Gaussian
function. We recall here that the dispersion around the \mz\ relation
was just $\sim$0.07 dex, a markedly small error taking into account
the nominal errors of the parameters involved. In summary, we cannot
confirm the existence of a dependency of the \mz\ relation on the
SFR, with the current CALIFA data.

{ A possible concern regarding this result may be that the limited
  number of objects in the current CALIFA sample may hamper the
  detection of a significant correlation.  We perform a simple
  Monte-Carlo simulation to test this possibility. For doing so, we
  assume that the \mz\ -SFR distribution follows the analytical
  formulae described by \cite{mann10}. Then, we extract a random
  subsample of 100 objects covering a similar range of the 
  parameters covered by the CALIFA sample, and we add a gaussian noise of 0.05
  dex \cite[i.e., the value reported by ][]{mann10}. For this
  subsample, we subtracted the \mz\ relation predicted by the same
  authors, and analyze the possible correlation between the residual
  of the abundance and the star-formation rate.  After 1000 realizations we found that
  in all the cases we found a clear correlation ($r\sim0.7$), with an
  slope compatible with the one predicted by the FMR relation. Thus, the
  number of objects does not affect the result. }

\subsection{Dependence of the \Sz\ relation on the SFR}\label{local}

In \cite{rosales12} we presented a new scaling relation between the
oxygen abundance and the surface mass density for individual
\HII regions (\Sz\ relation)
\footnote{An online annimated version of this relation
for the current analysed data is available at \url{http://www.youtube.com/watch?v=F7JCX-d7uPY}}.This relation is a local version of the global
\mz\ relation, and indeed, it is possible to recover this latter one from
the former, just by integrating over a certain aperture
\citep{rosales12}. We proposed that this relation is the real
fundamental one and the \mz\ relation being a natural consequence.  Following this
reasoning, it is important to test if there is a secondary relation
between the local-\mz\ relation and the SFR. An additional advantage of
this exploration is the increase of statistical significance of the
results, due to the much larger number of sampled objects here ($\sim$3000
\HII regions).

Figure \ref{fig:MZ_local}, top-left panel, shows the
\Sz\ relation for the \HII\ regions of the CALIFA galaxies
explored so far. As a first result, we reproduce the relation described
in \cite{rosales12}, using a completely different dataset \cite[the
  catalog of \HII regions extracted from 38 face-on spirals, using IFS
  data, by][]{sanchez12b}. Both quantities have a strong correlation,
with a correlation coefficient of $r$=0.98. As in the case of the
global \mz\ relation, we adopted the asymptotic equation \ref{eq1}.
In this case, the $x$ parameter represents the logarithm of the
stellar mass surface density in units of M$_{\odot}$pc$^{-2}$. The
$a$, $b$ and $c$ parameters have the same interpretation as those 
derived for the global relation, and we derive similar values when
fitting the data: $a=$8.86$\pm$0.01, $b$=0.97$\pm$0.07 and
$c$=2.80$\pm$0.21, reinforcing our hypothesis that both relations are
strongly linked. In general, the global-\mz\ relation has the
same shape as the local one but shifted arbitrarily in mass (to match them to the corresponding mass densities).

Like in the case of the integrated properties of the galaxies, there
is a tight correlation between the stellar surface mass density and
the surface star-formation density (Fig. \ref{fig:MZ_local}, top-right
panel). The correlation is a clear correlation ($r=0.66$), and it is well
described by a simple linear regression. The surface star-formation
density scales with the surface mass density with a slope of
$\alpha$=0.66$\pm$0.18, once more, below the expected value derived in
semi-analytical simulations for the Mass-SFR relation.  Note that a
comparison with a simulation taking into account the described local
relationship is not feasible at present, since no such simulation
exists  to our knowledge. 

{ This relation indicates that the HII regions with stronger
  star-formation are located in more massive areas of the galaxies
  (i.e., towards the center).  We have explored if our result is
  affected by the presence of young stars that biases our mass
  derivation. However, we found that our surface-brightness and colors
  (the two parameters used to derive the masses), are not strongly
  correlated with either the age of the underlying stellar population
  or the fraction of young stars. Evenmore, the SFR-density presents a
  positive weak trend with the B-V color, not the other way around.}

Like in the global case, these two relations between the oxygen abundance and the
surface star-formation rate with the surface mass density, may induce a
third relation between the two former parameters. In fact, there is a
trend between both parameters, much weaker than the other two
described here ($r=0.59$), as expected from an induced
correlation. Figure \ref{fig:MZ_local} bottom left panel illustrates
this dependence, showing the local-\mz\ relation, color-coded by the
mean surface star-formation rate per bin of oxygen abundance and
surface mass density. There is clear gradient in the surface SFR which
mostly follows the local-\mz\ relation. In this figure we
can see a deviation from this global trend in agreement with the
results by \cite{lara10a} and \cite{mann10}, in the sense that there
are a few \HII\ regions with low metal content and high star-formation
rate for their corresponding mass surface density (bottom right edge
of the figure). However, all together they represent less than a
2-3\%\ of the total sampled regions, and therefore cannot explain a
global effect as the one described in those publications. Furthermore,
the stochastic fluctuations around the average values at this low
number statistics may clearly produce this effect, since there are
other \HII\ regions in the same location of the diagram with much
lower surface SFR.

Following the same procedure outlined in Sec. \ref{mz_sfr}, we
removed the dependence of the oxygen abundance on the surface mass
density by subtracting the derived relation between both parameters.
Then, we explore possible dependencies on the surface density SFR. Figure
\ref{fig:MZ_local}, bottom-right panel, illustrates the result of this
analysis. It shows the distribution of the differential oxygen
abundance along the surface star-formation rate for the individual
\HII\ regions. No correlation is found between both for the individual
values ($r=0.17$). However, the dispersion around the zero-value is
larger than that reported for the similar residual distribution for the
global relation ($\sigma_{\rm local}=$0.11 dex). The distribution
seems to show some tail towards higher abundances for weaker SFRs and
lower for stronger ones. Thus, in order to explore possible
dependencies even more, we derive average differential oxygen
abundance for equal ranges of surface SFR of 0.2 dex width, and
determined the correlation coefficient. In this case we found a
clearer trend ($r=0.67$), in the sense that \HII\ regions with higher
surface SFR shows slightly lower abundances for similar
masses. However, the slope of the derived regression
$\alpha=-$0.0155$\pm$0.217 is compatible with no trend at all, and it
is far below the reported value of $\alpha\sim -$0.3 for the
dependence of the \mz\ relation and the SFR \cite{mann10} or
$\alpha\sim -$0.4 by \cite{lara10a}. Applying any of these
correlations (i.e., adding $\alpha\cdot$log(SFR) to the mass) does not
reduce the dispersion found around the zero value at all
($\sigma_{local,SFR}=$0.108 dex). In summary, we do not find a
correlation between the local-\mz\ relation with the SFR, in the way
described by other authors for the global one. Our results for the
local and global \mz\ relations are therefore fully consistent.

\begin{figure}
\centering
\includegraphics[width=6cm,angle=270]{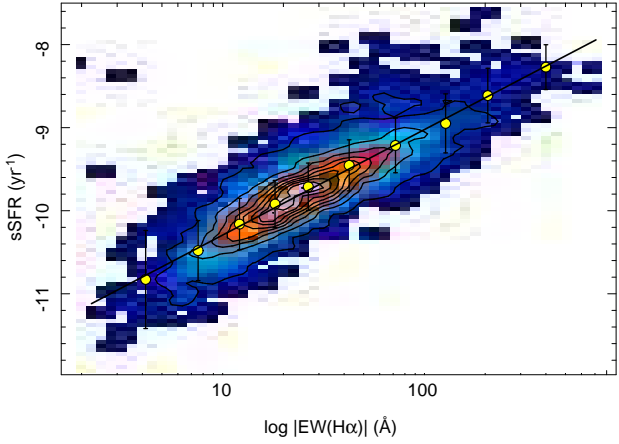}
\caption{\label{fig:EW} Distribution of the specific star formation as a function of the absolute equivalent width of
H$\alpha$, for the individual \HII\ regions. The solid yellow circles indicate the location of the average values in bins of 0.2 dex of the equivalent width of H$\alpha$. The solid line shows the best fitted linear regression line}
\end{figure}

\subsection{Dependence of the \mz\ relation on the sSFR}\label{sSFR}

Following \cite{mann10}, we have explored the possible dependence of
the \mz\ relation on the specific star-formation rate (sSFR), a
measurement of the current star-formation activity of the galaxy in
comparison with the past one. It is known to exhibit a smooth
dependence on the stellar mass (sSFR$\propto$M$^{-1}$), which has
been described at different redshifts \citep[e.g.][]{elbaz07}. Many
properties of galaxies depend on this parameter. In particular,
\cite{mann10} found that there was an apparent threshold in the sSFR
at about 10$^{-10}$ yr$^{-1}$ which discriminates the effect of the
SFR on the \mz\ relation, in the sense that the depence is stronger
for higher values of the sSFR. In this section we explore the possible
dependence of the \mz\ on sSFR.

Figure \ref{fig:MZ_sSFR} shows both the global and local
\mz\ relations for the galaxies and \HII\ regions discussed in this
article. In this case the color code indicates the sSFR, determined by
dividing the SFR and stellar masses described before. In both cases
the sSFR seems to follow the dependence of the abundance with the mass,
without a significant deviation from the trend along this
relation. Thus, we do not identify neither a clear secondary dependence on
the sSFR, nor a clear threshold at which the global and local
\mz\ relation show a change in the trend or an offset with respect to
the average relation. 

In the case of the \Sz\ relation, the data seem to follow a smooth
surface in 3D space, with the three parameters clearly correlated.
This surface was already discussed by \cite{rosales12}, who, instead
of using the sSFR used the equivalent width of H$\alpha$, a
well-known proxy for the former parameter
\citep[e.g.][]{kennicutt98}. Figure \ref{fig:EW} shows the empirical
relation for both parameters based on the individual
\HII\ regions. The correlation is very strong ($r=0.84$), and well
described with a simple linear regression, with a dispersion 0.26 dex:

\begin{equation}
 {\rm log}({\rm sSFR/yr}) = -11.38_{\pm 0.01} + 1.23_{\pm 0.02}
\cdot {\rm log|EW(H\alpha)/\AA|}
\end{equation}

\noindent Adopting this relation, the described curve is basically the
same as discussed by \cite{rosales12}, considering the
reparametrization of one of the axes. We note here that it is required
to take the absolute value of the equivalent width of H$\alpha$, which
has always negative values for strong emission line regions.

\section{Discussion and conclusions}\label{summary}

Along this study we have explored the global and local \mz\ relations
and their possible dependence on the SFR and sSFR based on the ionized
gas properties of the first 150 galaxies observed by the CALIFA
survey. So far, we have not found any dependence on the SFR and sSFR
different than that induced by the well known relation of both
parameters with the stellar mass. In addition, we explored similar
dependencies for the sample of $\sim$3000 individual \HII\ regions extracted
from the IFS data. We confirm the \Sz\ relation
described in \cite{rosales12}. However, no secondary dependence is
found between this relation and the SFR (or sSFR), different than that induced by their dependences on the stellar mass.

Our results contradict those of \cite{mann10} and \cite{lara10a}, who
described a trend for which galaxies with stronger star-formation show
lower oxygen abundances for the same mass range. Similar results were
already presented by \cite{hugh12}, using a sample of galaxies
observed using drift-scan techniques. The main difference between both
datasets is that while \cite{mann10} and \cite{lara10a} use the
spectroscopic data provided by the SDSS (i.e., single aperture, with a
strong aperture bias), both \cite{hugh12} and the results presented
here are based on integrated (and spatially resolved) properties of
the galaxies. { \cite{shen03} show that the physical size of
  galaxies in the SDSS survey comprises a range of values between a few kpcs
  to a few tens of kpcs. This means that the SDSS fiber covers a range
  between { $\sim$0.3 and $\sim$10 times the effective radii} of galaxies at
  the redshift range covered by \cite{mann10} and \cite{lara10a},
  depending both on their intrinsic properties and the involved
  cosmological distances. This aperture bias affects mostly the
  derived SFR \cite[depending on the aperture corretion applied, which
    are under debate, e.g.][]{gers12} and the oxygen abundances
  \cite[e.g.][]{sanchez12a,sanchez12b}, but in some cases it may also
  affect the mass derivation (if the spectroscopic information is used
  to derive the Mass-to-Light ratio).}


In addition, the redshift ranges covered by these studies are also
different. The data presented by \cite{hugh12} and ours are focused on
relatively nearby galaxies ($D<120$Mpc), with a narrow range of
distances in both cases. In contrast, SDSS data used by \cite{mann10}
and \cite{lara10a} also differ ($0.07<z<0.30$ and $0.04<z<0.1$,
respectively), however they both correspond to a much wider redshift
range than that covered by us. In both cases the aperture biases of
the SDSS spectra have two effects: (i) They grab a larger portion of
the galaxies at higher-redshifts, for the same masses, which affects
their estimations of the abundance (decreasing the value at higher
redshifts) and the SFR (increasing the value at higher redshifts too);
(ii) Due to the reported redshift evolution of the \mz\ relation, they
derive a lower abundance for higher redshift galaxies than for
lower-redshift ones, for the same mass, even in the absence of any
aperture bias. { This second effect may be less important, since it would be at maximum of the order
  of $\sim$0.06 dex for \cite{mann10}, but just $\sim$0.02 dex for
  \cite{lara10a}.  The first effect is also stronger for the case of
  the study by \cite{mann10}}. We can reproduce qualitatively the
secondary trend with the SFR of the \mz\ relation by taking our IFS
data and simulate different SDSS aperture spectra corresponding to
different projected distances (see Appendix \ref{ape}). However, the
derived slope is not reliable due to the uncertainties involved
(different sample selections, different methods to derive the
metallicity, unclear evolution of \mz\ relation with the
redshift). Although we cannot unambiguously prove that aperture
effects are inducing this relation, qualitatively there is certainly
an agreement.

As described by \cite{mann10}, oxygen abundance is a rather simple parameter, which is governed by three
processes: (i) star-formation, which transforms gas into stellar
mass, increasing the metal content; (ii) gas accretion, or inflows,
which dilutes the oxygen abundance (assuming accretion of more or less
{\it pristine} gas); and (iii) gas outflows, induced by the
star-formation. However, none of these processes is simple, and their
interdependence is unclear. 

Based on the Schmidt-Kennicutt law, the surface SFR depends on the
surface mass density, $\mu {\rm SFR} \propto {\Sigma_{gas}}^n$
\citep{kennicutt98}, with a slope larger than one ($n\sim
1.4-1.5$). This means that the integrated SFR of a galaxy should
depend on the gas inflow rate \citep[e.g.][]{mann10}. However, this
does not necessarily mean a global dilution and a decrease in the
average oxygen abundance, since inflow is a radial process while
star-formation is a local one. Gas that falls into the inner regions
may be polluted by star formation further out if the latter process is
fast enough.  Therefore, if the typical timescale to recycle gas is
shorter than the inflow timescale \cite[e.g.][]{silk93}, no SFR
dependence is expected. { On the other hand, if the timescales for
  the feedback and global gas recycling are much longer than the
  typical length of a star-formation episode \cite[$\sim$1 Gyr for
    L$^*$ galaxies][]{quil08}, there will be a lose of connection
  between the starformation and the global metal enrichment, too.}


On the other hand, outflows induced by strong star-formation and the
subsequent winds due to SN explosions are known to eject metals into
the intergalactic medium, and therefore regulate the metal content of
galaxies \citep[e.g.][]{lars74,priet08,pala05}. It is therefore expected that the strength of
the outflows, and the fraction of metals lost, depends on the SFR. However, the detailed relation is difficult to address, since
the ability of an outflow to eject metals could be counter-balanced by
the gravitational field of the galaxy, which makes metals to fall back
into the disk in a back rain process. The high metallicity of the diffuse IGM
in galaxy groups and clusters is indirect evidence that at
least some fraction of the metals in these winds is lost to
the environment \citep[e.g.][]{renz93}.

 Although it is hard to study just using the
integrated properties of galaxies, the spatial distribution of metals
and, in particular, the dispersion of the oxygen abundance at a
certain distance from the center could be used to constrain the
strength of these (and other) diffusion processes. Our preliminary
results, \cite{sanchez12b}, indicate that this dispersion is rather
low. Therefore, either metals are really expelled and lost and/or the
net effect is not very strong. The lack of a clear secondary
dependence of the oxygen abundance, both globally or locally on the
SFR and the $\mu${SFR} also suggests that this is not the dominant
process.

Following the formalism presented by \cite{mann10}, the fraction of
metals ejected due to outflows has to be compensated by the metals
produced by the star-formation process. If, instead of the reported
dependence of the abundance on the SFR of $\sim -$0.32$\cdot$log(SFR), we
consider no dependence at all, and using their formulae, the
amount of metals lost by this process should be $\Delta {\rm
  log(O/H)}\propto$ SFR$^{0.33}$, i.e., much smaller than that required
to reproduce their results.

In summary our results are consistent with those already presented
in \cite{sanchez12b} and \cite{rosales12}. In general, the properties
of the ionized gas in late-type galaxies are consistent with a
quiescent evolution, where gas recycling is faster than other times
scales involved \citep[][]{silk93}. This would imply that the galaxies
seem to behave locally in a similar manner than globally, dominated by a radial mass distribution following the potential well of the matter, with an
inside-out growth that is regulated by gas inflow and local
down-sizing star-formation. Therefore, the dominant parameter that
defines the amount of metals is the stellar mass, since both
parameters are the consequence of an (almost) closed-box star-formation
process. However, we should remark that our results are only valid 
for galaxies with stellar masses higher than $\sim$10$^{9.5}$ M$_\odot$,
where the CALIFA sample becomes complete.

\input{Mass_OH.tab.csv}


\begin{acknowledgements}

We thank the director of CEFCA, Dr. M. Moles, for his sincere support of this project.

This study makes uses of the data provided by the Calar Alto Legacy
Integral Field Area (CALIFA) survey (http://califa.caha.es/).

CALIFA is the first legacy survey being performed at Calar Alto. The
CALIFA collaboration would like to thank the IAA-CSIC and MPIA-MPG as
major partners of the observatory, and CAHA itself, for the unique
access to telescope time and support in manpower and infrastructures.
The CALIFA collaboration thanks also the CAHA staff for the dedication
to this project.

Based on observations collected at the Centro Astron\'omico Hispano
Alem\'an (CAHA) at Calar Alto, operated jointly by the
Max-Planck-Institut f\"ur Astronomie and the Instituto de Astrof\'\i sica de
Andalucia (CSIC).

We thank the {\it Viabilidad , Dise\~no , Acceso y Mejora } funding program,
ICTS-2009-10, for supporting the initial developement of this project.

S.F.S., F.F.R.O. and D. Mast thank the {\it Plan Nacional de Investigaci\'on y Desarrollo} funding programs, AYA2010-22111-C03-03 and AYA2010-10904E, of the Spanish {\it Ministerio de   Ciencia e Innovaci\'on}, for the support given to this project. 

S.F.S thanks the the {\it Ram\'on y Cajal} project RyC-2011-07590 of the spanish {\it Ministerio de Econom\'\i a y Competitividad}, for the support giving to this project.

SFS and BJ acknowledge suuport  by the grants No. M100031241 and
M100031201 of the Academy of Sciences of the Czech Republic
(ASCR Internal support program of international cooperation projects -
PIPPMS)
and by the Czech Republic program for the long-term
development of the research institution No. RVO67985815.

R.G.D , E.P. and R.G.B. thank the {\it Plan Nacional de Investigaci\'on y Desarrollo} funding program AYA2010-15081.

F.F.R.O. acknowledges the Mexican National Council for Science and
Technology (CONACYT) for financial support under the programme
Estancias Posdoctorales y Sab\'aticas al Extranjero para la
Consolidaci\'on de Grupos de Investigaci\'on, 2010-2011

J.M. and J.P. acknowledge financial support from the Spanish grant
AYA2010-15169 and Junta de Andaluc\'{\i}a TIC114 and Excellence Project P08-TIC-03531.

D. M. and A. M.-I. are supported by the Spanish Research Council within
the program JAE-Doc, Junta para la Ampliaci\'on de Estudios, co-funded by
the FSE.

R.A. Marino was also funded by the spanish programme of International Campus of Excellence Moncloa (CEI).

J. I.-P., J. M. V., A. M.-I. and C. K. have been partially funded by the projects AYA2010-21887 from the Spanish PNAYA,  CSD2006 - 00070  ``1st Science with
GTC''  from the CONSOLIDER 2010 programme of the Spanish MICINN, and TIC114 Galaxias y Cosmolog\'{\i}a of the Junta de Andaluc\'{\i}a (Spain). 

M.A.P.T. acknowledges support by the Spanish MICINN through grant AYA2012-38491-C02-02, and by the Autonomic Government of Andalusia through grants P08-TIC-4075 and TIC-126.

Polychronis Papaderos is supported by a Ciencia 2008 contract,
funded by FCT/MCTES (Portugal) and POPH/FSE (EC).

Jean Michel Gomes is supported by grant SFRH/BPD/66958/2009 from FCT (Portugal).

This paper makes use of the Sloan Digital Sky Survey data. Funding for the
SDSS and SDSS-II has been provided by the Alfred P. Sloan Foundation,  the
Participating Institutions,  the National Science Foundation,  the
U.S. Department of Energy,  the National Aeronautics and Space Administration, 
the Japanese Monbukagakusho,  the Max Planck Society,  and the Higher Education
Funding Council for England. The SDSS Web Site is http://www.sdss.org/.

The SDSS is managed by the Astrophysical Research Consortium for the
Participating Institutions. The Participating Institutions are the
American Museum of Natural History,  Astrophysical Institute Potsdam, 
University of Basel,  University of Cambridge,  Case Western Reserve
University,  University of Chicago,  Drexel University,  Fermilab,  the
Institute for Advanced Study,  the Japan Participation Group,  Johns Hopkins
University,  the Joint Institute for Nuclear Astrophysics,  the Kavli
Institute for Particle Astrophysics and Cosmology,  the Korean Scientist
Group,  the Chinese Academy of Sciences (LAMOST),  Los Alamos National
Laboratory,  the Max-Planck-Institute for Astronomy (MPIA),  the
Max-Planck-Institute for Astrophysics (MPA),  New Mexico State University, 
Ohio State University,  University of Pittsburgh,  University of Portsmouth, 
Princeton University,  the United States Naval Observatory,  and the
University of Washington.

This publication makes use of data products from the Two Micron All
Sky Survey, which is a joint project of the University of
Massachusetts and the Infrared Processing and Analysis
Center/California Institute of Technology, funded by the National
Aeronautics and Space Administration and the National Science
Foundation.

\end{acknowledgements}

\bibliography{CALIFAI}
\bibliographystyle{aa}

\appendix

\section{Dependence of the \mz\ relation with the SFR: Apperture effects.}
\label{ape}

\begin{figure*}
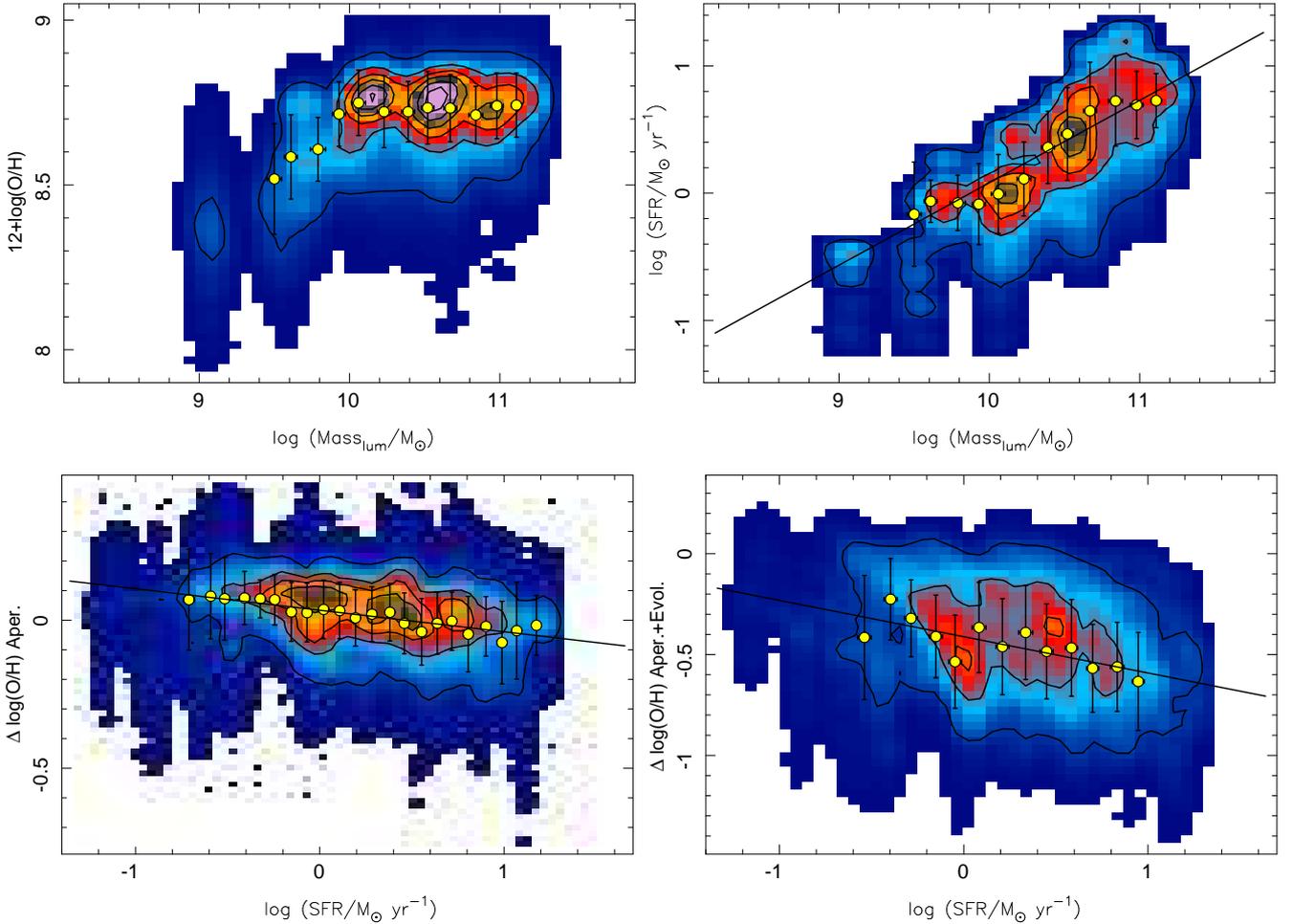

\centering
\includegraphics[width=6.5cm,angle=270,clip,trim=0 0 -20 0]{figs/SIM_M_Z.ps}
\includegraphics[width=6.5cm,angle=270,clip,trim=0 0 -20 0]{figs/SIM_M_SFR.ps}
\includegraphics[width=6.5cm,angle=270,clip,trim=0 0 -20 0]{figs/SIM_SFR_DELTA_OH_1.ps}
\includegraphics[width=6.5cm,angle=270,clip,trim=0 0 -20 0]{figs/SIM_SFR_DELTA_OH_2.ps}
\caption{\label{fig:ape} {\it Top-Right panel:} Distribution of the oxygen abundance along the stellar mass for the $\sim$50,000 simulated aperture measurements, as described in the text. The contours indicate the same encircled fraction of object than in precendent figures (e.g., Fig. \ref{fig:MZ_local}). The solid yellow circles indicate the average abundances for consecutive bins of 0.2 dex in stellar mass. Only those binned values with a significant number of objects are shown. {\it Top-left panel:} Distribution of the integrated star-formation rate along the stellar mass for the simulated galaxies. Contours and symbols represent the same as in previous panel. The solid line shows the regression line between the two parameters. {\it Bottom-left panel:} Distribution of the differential oxygen abundance with respect to the derived \mz\ relation along the SFR for the simulated galaxies, when only the aperture effect is considered. {\it Bottom-right panel:} Distribution of the differential oxygen abundance with respect to the derived \mz\ relation along the SFR for the simulated galaxies, when both the aperture effect and the evolution of the \mz\ relation with redshift is taken into account. In both panels the contours and sybmols represent similar concepts like in previous figures and panels. The solid line shows the best derived fit using a simple linear regression method.}
\end{figure*}

We show in the main text our results on the study of the possible
dependence of the \mz\ relation with the SFR. Using the IFS data
provided by the CALIFA survey, we were able to derive the
\mz\ relation with abundances not affected by aperture effects.  In a
similar way we derive the complete integrated SFR for these
galaxies. Contrary to previous results in the literature
\cite[e.g.][]{mann10,lara10a} we find no secondary dependence of the
abundance with the SFR once considered the dependence of both
parameters with the Mass. The only similar result found in the
literature where no dependence on the SFR is described used a sample
of integrated spectra taken using drift-scan observations
\citep[e.g.][]{hugh12}. In summary, when the biases due to aperture
are minimized the dependence on the SFR seems to disappear.

It is beyond the scope of this article to explain why other studies
find a correlation that we cannot reproduce, in particular when we use
a less biased and more detailed estimation of the considered
quantities.  However, as a sanity check, and in order to provide a
suitable explanation, we performed a test of the aperture effects. The
oxygen abundance and the SFR are potentially affected by this aperture
effect. However, the stellar mass is the less affected, in particular
if only photometric data is considered.  With CALIFA we are in the
unique situation of being able to reproduce the aperture effects
present in single aperture observations, by simulating the considered
apertures and re-analysing the corresponding {\it simulated} spectra.

{ It is important to use as our starting dataset for the simulations
a sample that we know {\it a priori} that it does not show any secondary correlation
between the abundance and the SFR, when the dependency with 
the Mass is removed. In this case we will be sure that the induced correlation
was not present as an input ingredient of the simulation.  }

For each galaxy in our dataset with well defined characteristic oxygen
abundance, we extracted a set of aperture spectra centred in the peak
emission of the galaxy and with consecutive apertures corresponding to
the distance of each of the detected \HII\ regions from our catalogue.
This ensures that there is a clear change in the ionization conditions
between consecutive apertures. Each of these aperture spectra correspond
to the {\it same} galaxy observed at different redshifts. To be
consistent with the redshift ranges adopted by \cite{mann10} and \cite{lara10a}, we
restricted our apertures to diameters larger than 12$\arcsec$.  This
roughly corresponds to the projected size of a SDSS fiber at the lower
redshift adopted by this study ($z\sim$0.07). Once extracted the
spectra, we repeat the same analysis that we perform for the individual
\HII\ regions, described in Sec. \ref{ana_par}, deriving the oxygen
abundance and integrated SFR. Together with the integrated stellar
masses, we ended up with $\sim$2500 individual simulations of the three
quantities analysed in this study.

Finally, we perform a Monte-Carlo simulation to take into account the
possible effects of the noise in the overall process. For each of the
three considered quantities 20 individual realizations were derived,
taking into account the corresponding propagated errors (including
also those of the stellar masses). We end up with $\sim$50,000
simulated {\it aperture biased} quantities.

Figure \ref{fig:ape} shows the results of this simulation. In the
top-left panel we show the derived \mz\ relation for the simulated
spectra. As expected it follows the observed relation, presented in
Fig. \ref{fig:MZ}. { This result was already shown by \cite{rosales12},
where they reproduce the global \mz\ from the local one (derived from the
abundance gradients), based purely on simulated galaxies.} However, a more detailed comparison shows that
although both relations are very similar there are slightly
differences: (i) the asymptotic oxygen abundance for large masses is
$\sim$0.05 dex larger in the simulated relation and (ii) the
dispersion of oxygen abundances along the average value is larger
($\sigma_{\rm log (O/H)}\sim$0.1 dex, similar to the one reported by
T04). Both results are expected: (1) we are including more central
estimations of the abundance for the same range of masses. In the
center of the galaxies the abundance is larger than in the outer parts
by $\sim$0.1 dex \citep[e.g.][]{sanchez12b}. Thus, the net effect is
an increase of the upper envelop of the abundance; (2) we are
including measurements of different abundances for the same mass range
as we enlarge the covered area of the galaxy, simulating the effect of
the redshift. Therefore, the dispersion at a certain mass range is
increased.

Figure \ref{fig:ape}, top-righ panel shows the distribution of the
simulated SFRs along the stellar masses. As expected we reproduce the
relation between both quantities, in a similar way as described in
Fig. \ref{fig:MZ}. The main differences are (1) the dispersion is
slightly larger: $\sigma_{\rm log (SFR)_{\sim}} \sim$0.35 dex,
compared to $\sigma_{\rm log (SFR)} \sim$0.18 dex; (2) there
is a tail to lower values of the SFR at a given mass. This is in
particular evident for the low-mass range of the figure (${\rm M}\sim
10^{9.5}{\rm M}\odot$). Once more, these effects are clearly explained
by the aperture bias, since at a given mass we are {\it sampling}
areas of lower integrated SFR than the real one (i.e., the central
regions), which is an upper envelop. 

Finally, we derive the differential oxygen abundance with
respect to the considered \mz\ relation, for the simulated data. We
try to determine if there is any correlation with the SFR, following
the analysis in Sec. \ref{mz_sfr}. Figure \ref{fig:ape},
bottom-left panel shows the distribution of these two
parameters. There is a clear correlation ($r=0.92$), in the similar
sense as described by \cite{mann10} and \cite{lara10a}, i.e., galaxies
with lower SFR seem to be more metal poor. Once applied this
correlation, we recover the initial dispersion of the \mz\ relation
described for our original dataset ($\sigma \Delta {\rm
  log(O/H)}\sim$0.07 dex), a decrease similar to the one
reported by these two authors. { The dependence of the 
SFR is $\Delta_{\rm log(O/H)}\propto$$-$0.1 log(SFR), just slightly weaker than the
linear term of the equation 2 of \cite{mann10}, $-$0.14 dex/log(SFR), or the
combination of equations 3 and 4 of the same authors. }

In order to investigate further this issue, we have included in our
simulation the reported \mz\ evolution with redshift, as described by
\cite{mous11}. For doing so, we have determined at which redshift
corresponds each of the simulated apertures, and we apply the
described correction to the corresponding derived abundance.
Results are shown in Figure \ref{fig:ape}, bottom-right panel.
Taking into account the possible redshift dependence, the new differential 
oxygen abundance have a sharper dependence on the SFR, 
$\Delta {\rm log(O/H)}$$\propto$$-$0.2log(SFR).

Despite of the fact that this is a very simplistic simulation, based on a reduced
number of input galaxies, in general we have shown that aperture effects,
together with other evolutionary properties of galaxies, may induce the 
secondary correlations between the \mz\ and the SFR that we are not able to
reproduce using our IFS data.

\end{document}

%% file: Mass_OH.tab.csv.tex
\begin{table*}
 \caption{Properties of the sample of galaxies.}
 \label{table_Mass}
 \begin{center}
 \begin{tabular}{lrrrrrrrr}
 \hline\hline 
Galaxy &  redshift & V-band (mag)& B-V & Type & N$_{\rm HII}$ & log(Mass/M$\odot$) & 12+log(O/H) & log(SFR/M$\odot$ yr$^{-1}$)\\
\hline  
IC5376            & 0.01663& 13.84 $\pm$  0.02 &  0.89 $\pm$  0.03& SbA  &   2& 10.55 $\pm$  0.19&   \nodata&   \nodata \\
NGC7819           & 0.01730& 13.80 $\pm$  0.02 &  0.71 $\pm$  0.03& ScA  &  31& 10.41 $\pm$  0.17&   8.63 $\pm$   0.07&   0.48 $\pm$   0.01 \\
UGC00036          & 0.02089& 13.72 $\pm$  0.01 &  0.95 $\pm$  0.01& SabAB&   5& 10.88 $\pm$  0.20&   8.72 $\pm$   0.04&   0.40 $\pm$   0.01 \\
NGC0001           & 0.01502& 13.11 $\pm$  0.01 &  0.85 $\pm$  0.01& SbcA &  31& 10.75 $\pm$  0.19&   8.75 $\pm$   0.06&   0.93 $\pm$   0.01 \\
NGC0036           & 0.01993& 13.06 $\pm$  0.01 &  0.93 $\pm$  0.01& SbB  &  25& 11.08 $\pm$  0.20&   8.71 $\pm$   0.04&   0.76 $\pm$   0.01 \\
UGC00312          & 0.01532& 13.73 $\pm$  0.02 &  0.48 $\pm$  0.02& SdB  &   0& 10.06 $\pm$  0.14&   \nodata&   \nodata \\
NGC0444           & 0.01596& 14.34 $\pm$  0.03 &  0.63 $\pm$  0.04& ScdA &  11&  9.98 $\pm$  0.16&   8.52 $\pm$   0.06&   0.10 $\pm$   0.01 \\
NGC0477           & 0.01989& 14.05 $\pm$  0.02 &  0.81 $\pm$  0.02& SbcAB&  24& 10.53 $\pm$  0.18&   8.64 $\pm$   0.11&   0.75 $\pm$   0.01 \\
IC1683            & 0.01584& 13.67 $\pm$  0.01 &  0.88 $\pm$  0.02& SbAB &   4& 10.55 $\pm$  0.19&   8.74 $\pm$   0.03&   0.77 $\pm$   0.01 \\
NGC0496           & 0.02003& 13.71 $\pm$  0.02 &  0.68 $\pm$  0.02& ScdA &  30& 10.57 $\pm$  0.17&   8.62 $\pm$   0.07&   0.81 $\pm$   0.01 \\
UGC01057          & 0.02107& 14.23 $\pm$  0.02 &  0.62 $\pm$  0.03& ScAB &  21& 10.33 $\pm$  0.16&   8.55 $\pm$   0.15&   0.60 $\pm$   0.01 \\
NGC0776           & 0.01616& 13.00 $\pm$  0.01 &  0.90 $\pm$  0.01& SbB  &  34& 10.91 $\pm$  0.20&   8.77 $\pm$   0.07&   0.88 $\pm$   0.01 \\
UGC01938          & 0.02118& 14.23 $\pm$  0.03 &  0.72 $\pm$  0.04& SbcAB&  15& 10.44 $\pm$  0.17&   8.60 $\pm$   0.08&   0.57 $\pm$   0.01 \\
NGC1056           & 0.01512& 12.56 $\pm$  0.01 &  0.92 $\pm$  0.01& SaA  &  10& 10.06 $\pm$  0.20&   8.62 $\pm$   0.03&   0.18 $\pm$   0.01 \\
NGC1167           & 0.01640& 12.26 $\pm$  0.01 &  0.98 $\pm$  0.01& S0A  &   1& 11.31 $\pm$  0.21&   \nodata&   \nodata \\
NGC1349           & 0.02191& 13.02 $\pm$  0.02 &  0.88 $\pm$  0.02& E6A  &   7& 11.13 $\pm$  0.19&   8.71 $\pm$   0.08&   0.35 $\pm$   0.02 \\
UGC03107          & 0.02773& 14.41 $\pm$  0.05 &  0.82 $\pm$  0.06& SbA  &   0& 10.71 $\pm$  0.19&   \nodata&   \nodata \\
UGC03253          & 0.01363& 13.30 $\pm$  0.01 &  0.71 $\pm$  0.02& SbB  &  26& 10.38 $\pm$  0.17&   8.70 $\pm$   0.06&   0.32 $\pm$   0.01 \\
NGC2253           & 0.01180& 12.65 $\pm$  0.02 &  0.79 $\pm$  0.02& SbcB &  20& 10.59 $\pm$  0.18&   8.76 $\pm$   0.03&   0.29 $\pm$   0.01 \\
NGC2410           & 0.01570& 12.94 $\pm$  0.01 &  0.86 $\pm$  0.01& SbAB &  19& 10.83 $\pm$  0.19&   8.65 $\pm$   0.05&   0.81 $\pm$   0.01 \\
UGC03944          & 0.01289& 13.92 $\pm$  0.03 &  0.57 $\pm$  0.03& SbcAB&  34&  9.88 $\pm$  0.16&   8.54 $\pm$   0.12&   0.03 $\pm$   0.01 \\
UGC03995          & 0.01601& 12.82 $\pm$  0.01 &  0.77 $\pm$  0.01& SbB  &  19& 10.84 $\pm$  0.18&   8.67 $\pm$   0.05&   0.51 $\pm$   0.01 \\
NGC2449           & 0.01629& 13.12 $\pm$  0.01 &  0.87 $\pm$  0.01& SabAB&   9& 10.83 $\pm$  0.19&   8.77 $\pm$   0.03&   0.33 $\pm$   0.01 \\
UGC04132          & 0.01718& 13.28 $\pm$  0.01 &  0.80 $\pm$  0.01& SbcAB&  11& 10.74 $\pm$  0.18&   8.67 $\pm$   0.06&   1.18 $\pm$   0.01 \\
UGC04461          & 0.01657& 13.87 $\pm$  0.02 &  0.55 $\pm$  0.03& SbcA &  19& 10.14 $\pm$  0.15&   8.48 $\pm$   0.10&   0.53 $\pm$   0.01 \\
IC2487            & 0.01427& 13.52 $\pm$  0.01 &  0.84 $\pm$  0.02& ScAB &  23& 10.51 $\pm$  0.19&   8.65 $\pm$   0.06&   0.44 $\pm$   0.01 \\
NGC2906           & 0.01706& 12.51 $\pm$  0.01 &  0.88 $\pm$  0.01& SbcA &  31& 10.34 $\pm$  0.19&   8.79 $\pm$   0.05&   0.17 $\pm$   0.01 \\
NGC2916           & 0.01229& 12.87 $\pm$  0.01 &  0.65 $\pm$  0.01& SbcA &  50& 10.41 $\pm$  0.16&   8.72 $\pm$   0.07&   0.47 $\pm$   0.01 \\
UGC05108          & 0.02692& 14.02 $\pm$  0.02 &  0.84 $\pm$  0.02& SbB  &   0& 10.87 $\pm$  0.19&   \nodata&   \nodata \\
UGC05358          & 0.01045& 14.99 $\pm$  0.03 &  0.64 $\pm$  0.04& SdB  &   6&  9.36 $\pm$  0.16&   8.41 $\pm$   0.05&  -0.43 $\pm$   0.02 \\
UGC05359          & 0.02812& 14.38 $\pm$  0.03 &  0.73 $\pm$  0.03& SbB  &   0& 10.65 $\pm$  0.17&   \nodata&   \nodata \\
UGC05396          & 0.01778& 14.14 $\pm$  0.02 &  0.67 $\pm$  0.03& SbcAB&  16& 10.23 $\pm$  0.17&   8.62 $\pm$   0.06&   0.39 $\pm$   0.01 \\
NGC3106           & 0.02074& 13.03 $\pm$  0.01 &  0.88 $\pm$  0.01& SabA &   9& 11.08 $\pm$  0.19&   8.71 $\pm$   0.07&   0.43 $\pm$   0.02 \\
NGC3057           & 0.01506& 13.86 $\pm$  0.03 &  0.43 $\pm$  0.03& SdmB &  22&  8.95 $\pm$  0.14&   8.34 $\pm$   0.08&  -0.49 $\pm$   0.01 \\
UGC05498NED01     & 0.02090& 14.21 $\pm$  0.02 &  0.94 $\pm$  0.03& SaA  &   0& 10.66 $\pm$  0.20&   \nodata&   \nodata \\
UGC05598          & 0.01875& 14.46 $\pm$  0.03 &  0.71 $\pm$  0.03& SbA  &  10& 10.19 $\pm$  0.17&   8.60 $\pm$   0.05&   0.44 $\pm$   0.01 \\
NGC3303           & 0.02124& 13.98 $\pm$  0.01 &  0.97 $\pm$  0.02& S0aAB&   0& 10.84 $\pm$  0.20&   \nodata&   \nodata \\
UGC05771          & 0.02473& 13.67 $\pm$  0.02 &  0.96 $\pm$  0.02& E6A  &   0& 11.08 $\pm$  0.20&   \nodata&   \nodata \\
NGC3381           & 0.01638& 13.03 $\pm$  0.01 &  0.54 $\pm$  0.01& SdB  &  49&  9.58 $\pm$  0.15&   8.64 $\pm$   0.06&  -0.05 $\pm$   0.01 \\
UGC06036          & 0.02168& 13.64 $\pm$  0.01 &  0.98 $\pm$  0.01& SaA  &   0& 10.99 $\pm$  0.21&   \nodata&   \nodata \\
UGC06312          & 0.02103& 13.81 $\pm$  0.02 &  0.93 $\pm$  0.02& SabA &   4& 10.87 $\pm$  0.20&   8.67 $\pm$   0.04&   0.49 $\pm$   0.01 \\
NGC3614           & 0.01762& 13.15 $\pm$  0.02 &  0.77 $\pm$  0.02& SbcAB&  52&  9.95 $\pm$  0.18&   8.71 $\pm$   0.09&   0.04 $\pm$   0.01 \\
NGC3687           & 0.01829& 12.80 $\pm$  0.01 &  0.65 $\pm$  0.01& SbB  &  53& 10.08 $\pm$  0.17&   8.70 $\pm$   0.05&  -0.09 $\pm$   0.01 \\
NGC3991           & 0.01171& 13.72 $\pm$  0.01 &  0.46 $\pm$  0.01& SmA  &   1&  9.75 $\pm$  0.14&   \nodata&   \nodata \\
NGC4003           & 0.02179& 13.60 $\pm$  0.01 &  0.93 $\pm$  0.02& S0aB &   0& 10.95 $\pm$  0.20&   \nodata&   \nodata \\
UGC07012          & 0.01030& 14.23 $\pm$  0.02 &  0.47 $\pm$  0.02& ScdAB&  13&  9.47 $\pm$  0.14&   8.49 $\pm$   0.08&  -0.06 $\pm$   0.01 \\
NGC4047           & 0.01125& 12.44 $\pm$  0.01 &  0.68 $\pm$  0.01& SbcA &  47& 10.55 $\pm$  0.17&   8.71 $\pm$   0.07&   0.71 $\pm$   0.01 \\
UGC07145          & 0.02195& 14.32 $\pm$  0.03 &  0.71 $\pm$  0.03& SbcA &  20& 10.40 $\pm$  0.17&   8.62 $\pm$   0.13&   0.76 $\pm$   0.01 \\
NGC4185           & 0.01284& 12.87 $\pm$  0.01 &  0.79 $\pm$  0.02& SbcAB&  30& 10.58 $\pm$  0.18&   8.75 $\pm$   0.06&   0.33 $\pm$   0.01 \\
NGC4210           & 0.01893& 12.92 $\pm$  0.01 &  0.72 $\pm$  0.01& SbB  &  56& 10.11 $\pm$  0.17&   8.73 $\pm$   0.06&   0.06 $\pm$   0.01 \\
\hline                    
 \end{tabular}
 \end{center}
\end{table*}

\addtocounter{table}{-1}

\begin{table*}
 \caption{Properties of the sample of galaxies.{\it Continue}}
 \label{table_Mass1}
 \begin{center}
 \begin{tabular}{lrrrrrrrr}
 \hline\hline 
Galaxy &  redshift & V-band (mag)& B-V & Type & N$_{\rm HII}$ & log(Mass/M$\odot$) & 12+log(O/H) & log(SFR/M$\odot$ yr$^{-1}$)\\
\hline  
IC0776            & 0.01922& 14.55 $\pm$  0.07 &  0.63 $\pm$  0.09& SdmA &  15&  9.43 $\pm$  0.17&   8.33 $\pm$   0.06&  -0.69 $\pm$   0.02 \\
NGC4470           & 0.01776& 12.66 $\pm$  0.01 &  0.52 $\pm$  0.01& ScA  &  18&  9.87 $\pm$  0.15&   8.59 $\pm$   0.03&   0.20 $\pm$   0.01 \\
NGC4676A          & 0.02242& 14.31 $\pm$  0.02 &  0.94 $\pm$  0.03&      &   0& 10.71 $\pm$  0.20&   \nodata&   \nodata \\
NGC4711           & 0.01353& 13.49 $\pm$  0.01 &  0.69 $\pm$  0.02& SbcA &   0& 10.27 $\pm$  0.17&   \nodata&   \nodata \\
UGC08107          & 0.02775& 13.85 $\pm$  0.02 &  0.89 $\pm$  0.02& SaA  &   0& 11.02 $\pm$  0.19&   \nodata&   \nodata \\
NGC4961           & 0.01845& 13.39 $\pm$  0.01 &  0.48 $\pm$  0.01& ScdB &  37&  9.63 $\pm$  0.14&   8.53 $\pm$   0.10&  -0.07 $\pm$   0.02 \\
NGC5000           & 0.01844& 13.55 $\pm$  0.02 &  0.74 $\pm$  0.02& SbcB &  27& 10.61 $\pm$  0.18&   8.74 $\pm$   0.05&   0.60 $\pm$   0.01 \\
UGC08250          & 0.01729& 14.93 $\pm$  0.04 &  0.63 $\pm$  0.05& ScA  &   6&  9.86 $\pm$  0.16&   8.50 $\pm$   0.06&   0.10 $\pm$   0.01 \\
UGC08267          & 0.02418& 14.44 $\pm$  0.03 &  0.91 $\pm$  0.03& SbAB &   2& 10.69 $\pm$  0.20&   \nodata&   \nodata \\
NGC5016           & 0.01848& 12.71 $\pm$  0.01 &  0.63 $\pm$  0.01& SbcA &  46& 10.09 $\pm$  0.16&   8.75 $\pm$   0.07&   0.18 $\pm$   0.01 \\
NGC5218           & 0.01974& 12.74 $\pm$  0.01 &  0.90 $\pm$  0.01& SabB &  12& 10.50 $\pm$  0.20&   8.70 $\pm$   0.05&   0.36 $\pm$   0.01 \\
UGC08733          & 0.01772& 14.29 $\pm$  0.03 &  0.68 $\pm$  0.03& SdmB &  12&  9.37 $\pm$  0.17&   8.40 $\pm$   0.09&  -0.47 $\pm$   0.01 \\
IC0944            & 0.02310& 13.45 $\pm$  0.01 &  0.92 $\pm$  0.02& SabA &   0& 11.08 $\pm$  0.20&   \nodata&   \nodata \\
UGC08778          & 0.01065& 13.84 $\pm$  0.01 &  0.78 $\pm$  0.02& SbA  &   3&  9.99 $\pm$  0.18&   8.67 $\pm$   0.01&  -0.44 $\pm$   0.01 \\
UGC08781          & 0.02518& 13.49 $\pm$  0.01 &  0.82 $\pm$  0.02& SbB  &  18& 11.02 $\pm$  0.19&   8.68 $\pm$   0.04&   0.61 $\pm$   0.01 \\
NGC5378           & 0.01977& 13.05 $\pm$  0.01 &  0.87 $\pm$  0.01& SbB  &   6& 10.33 $\pm$  0.19&   8.75 $\pm$   0.03&  -0.43 $\pm$   0.02 \\
NGC5406           & 0.01791& 12.73 $\pm$  0.01 &  0.84 $\pm$  0.01& SbB  &  49& 11.02 $\pm$  0.19&   8.78 $\pm$   0.06&   0.78 $\pm$   0.01 \\
UGC09067          & 0.03099& 13.96 $\pm$  0.02 &  0.66 $\pm$  0.02& SbcAB&   0& 10.81 $\pm$  0.17&   \nodata&   \nodata \\
NGC5614           & 0.01282& 12.16 $\pm$  0.01 &  0.95 $\pm$  0.01& SaA  &   8& 11.06 $\pm$  0.20&   8.72 $\pm$   0.03&   0.23 $\pm$   0.01 \\
NGC5633           & 0.01766& 12.61 $\pm$  0.01 &  0.65 $\pm$  0.01& SbcA &  29& 10.04 $\pm$  0.17&   8.77 $\pm$   0.06&   0.43 $\pm$   0.01 \\
NGC5630           & 0.01858& 13.39 $\pm$  0.01 &  0.46 $\pm$  0.01& SdmB &  14&  9.60 $\pm$  0.14&   8.39 $\pm$   0.05&   0.20 $\pm$   0.01 \\
UGC09291          & 0.01957& 13.58 $\pm$  0.02 &  0.61 $\pm$  0.02& ScdA &  27&  9.81 $\pm$  0.16&   8.60 $\pm$   0.10&  -0.04 $\pm$   0.01 \\
NGC5656           & 0.01051& 12.48 $\pm$  0.01 &  0.70 $\pm$  0.01& SbA  &  31& 10.46 $\pm$  0.17&   8.72 $\pm$   0.06&   0.49 $\pm$   0.01 \\
NGC5682           & 0.01724& 14.23 $\pm$  0.02 &  0.53 $\pm$  0.03& ScdB &   5&  9.22 $\pm$  0.15&   8.41 $\pm$   0.05&  -0.36 $\pm$   0.01 \\
NGC5720           & 0.02588& 13.68 $\pm$  0.01 &  0.76 $\pm$  0.02& SbcB &   6& 10.87 $\pm$  0.18&   8.61 $\pm$   0.06&   0.82 $\pm$   0.01 \\
NGC5732           & 0.01241& 13.83 $\pm$  0.01 &  0.57 $\pm$  0.02& SbcA &  24&  9.92 $\pm$  0.15&   8.63 $\pm$   0.12&   0.13 $\pm$   0.01 \\
UGC09476          & 0.01065& 13.33 $\pm$  0.01 &  0.66 $\pm$  0.02& SbcA &  46& 10.06 $\pm$  0.17&   8.68 $\pm$   0.05&   0.29 $\pm$   0.01 \\
NGC5735           & 0.01244& 13.37 $\pm$  0.02 &  0.75 $\pm$  0.02& SbcB &  52& 10.33 $\pm$  0.18&   8.65 $\pm$   0.10&   0.30 $\pm$   0.01 \\
NGC5772           & 0.01620& 12.88 $\pm$  0.01 &  0.82 $\pm$  0.01& SabA &  20& 10.87 $\pm$  0.19&   8.74 $\pm$   0.03&   0.42 $\pm$   0.01 \\
NGC5784           & 0.01815& 12.85 $\pm$  0.01 &  0.86 $\pm$  0.01& S0A  &   1& 11.03 $\pm$  0.19&   \nodata&   \nodata \\
UGC09598          & 0.01857& 13.94 $\pm$  0.02 &  0.74 $\pm$  0.03& SbcAB&  14& 10.45 $\pm$  0.18&   8.67 $\pm$   0.06&   0.27 $\pm$   0.01 \\
UGC09665          & 0.01838& 13.84 $\pm$  0.02 &  0.78 $\pm$  0.02& SbA  &   7&  9.80 $\pm$  0.18&   8.63 $\pm$   0.03&   0.03 $\pm$   0.01 \\
NGC5829           & 0.01865& 13.86 $\pm$  0.02 &  0.59 $\pm$  0.03& ScA  &  24& 10.31 $\pm$  0.16&   8.53 $\pm$   0.10&   0.55 $\pm$   0.02 \\
NGC5876           & 0.01093& 12.87 $\pm$  0.01 &  0.91 $\pm$  0.01& S0aB &   0& 10.55 $\pm$  0.20&   \nodata&   \nodata \\
NGC5888           & 0.02902& 13.36 $\pm$  0.01 &  0.86 $\pm$  0.02& SbB  &   0& 11.26 $\pm$  0.19&   \nodata&   \nodata \\
UGC09777          & 0.01547& 14.11 $\pm$  0.02 &  0.72 $\pm$  0.02& SbcA &   8& 10.19 $\pm$  0.17&   8.65 $\pm$   0.07&   0.33 $\pm$   0.01 \\
NGC5908           & 0.01094& 12.48 $\pm$  0.01 &  1.02 $\pm$  0.01& SaA  &   4& 10.83 $\pm$  0.21&   8.67 $\pm$   0.01&   0.53 $\pm$   0.01 \\
NGC5930           & 0.01861& 12.81 $\pm$  0.01 &  0.85 $\pm$  0.01& SabAB&   5& 10.31 $\pm$  0.19&   8.68 $\pm$   0.05&   0.32 $\pm$   0.01 \\
UGC09873          & 0.01844& 14.82 $\pm$  0.04 &  0.71 $\pm$  0.05& SbA  &   5& 10.05 $\pm$  0.17&   8.61 $\pm$   0.06&   0.35 $\pm$   0.01 \\
UGC09892          & 0.01886& 14.46 $\pm$  0.03 &  0.72 $\pm$  0.04& SbcA &   7& 10.20 $\pm$  0.17&   8.67 $\pm$   0.06&   0.25 $\pm$   0.01 \\
NGC5953           & 0.01671& 12.64 $\pm$  0.01 &  0.84 $\pm$  0.01& SaA  &   4& 10.10 $\pm$  0.19&   8.70 $\pm$   0.12&   0.10 $\pm$   0.01 \\
ARP220            & 0.01813& 13.31 $\pm$  0.01 &  0.77 $\pm$  0.02& SdA  &   0& 10.74 $\pm$  0.18&   \nodata&   \nodata \\
NGC5957           & 0.01600& 12.72 $\pm$  0.01 &  0.74 $\pm$  0.01& SbB  &  31&  9.95 $\pm$  0.18&   8.73 $\pm$   0.08&  -0.27 $\pm$   0.01 \\
IC4566            & 0.01846& 13.45 $\pm$  0.01 &  0.87 $\pm$  0.02& SbB  &   6& 10.80 $\pm$  0.19&   8.74 $\pm$   0.07&   0.38 $\pm$   0.01 \\
NGC5987           & 0.01992& 12.35 $\pm$  0.01 &  0.96 $\pm$  0.01& SaA  &   0& 10.73 $\pm$  0.20&   \nodata&   \nodata \\
UGC10043          & 0.01715& 14.74 $\pm$  0.07 &  0.91 $\pm$  0.09& SabAB&   4&  9.45 $\pm$  0.20&   8.58 $\pm$   0.03&  -0.66 $\pm$   0.01 \\
NGC6004           & 0.01272& 12.91 $\pm$  0.01 &  0.79 $\pm$  0.02& SbcB &  38& 10.56 $\pm$  0.18&   8.78 $\pm$   0.03&   0.52 $\pm$   0.01 \\
IC1151            & 0.01712& 13.25 $\pm$  0.01 &  0.52 $\pm$  0.01& ScdB &  22&  9.62 $\pm$  0.15&   8.50 $\pm$   0.06&  -0.02 $\pm$   0.01 \\
UGC10123          & 0.01239& 13.98 $\pm$  0.02 &  0.93 $\pm$  0.02& SabA &   3& 10.27 $\pm$  0.20&   8.68 $\pm$   0.03&   0.43 $\pm$   0.01 \\
NGC6032           & 0.01424& 13.58 $\pm$  0.02 &  0.83 $\pm$  0.02& SbcB &   2& 10.47 $\pm$  0.19&   \nodata&   \nodata \\
\hline                    
 \end{tabular}
 \end{center}
\end{table*}

\addtocounter{table}{-1}

\begin{table*}
 \caption{Properties of the sample of galaxies. {\it Continue}}
 \label{table_Mass2}
 \begin{center}
 \begin{tabular}{lrrrrrrrr}
 \hline\hline              
Galaxy &  redshift & V-band (mag)& B-V & Type & N$_{\rm HII}$ & log(Mass/M$\odot$) & 12+log(O/H) & log(SFR/M$\odot$ yr$^{-1}$)\\
\hline  
NGC6060           & 0.01446& 12.74 $\pm$  0.01 &  0.82 $\pm$  0.01& SbA  &  16& 10.81 $\pm$  0.19&   8.69 $\pm$   0.08&   0.86 $\pm$   0.01 \\
UGC10205          & 0.02185& 13.36 $\pm$  0.01 &  0.88 $\pm$  0.01& S0aA &   3& 10.99 $\pm$  0.19&   8.63 $\pm$   0.01&   0.52 $\pm$   0.01 \\
NGC6063           & 0.01938& 13.34 $\pm$  0.01 &  0.65 $\pm$  0.02& SbcA &  38&  9.96 $\pm$  0.16&   8.61 $\pm$   0.10&  -0.02 $\pm$   0.01 \\
IC1199            & 0.01557& 13.52 $\pm$  0.01 &  0.73 $\pm$  0.02& SbAB &  17& 10.45 $\pm$  0.18&   8.77 $\pm$   0.05&   0.42 $\pm$   0.01 \\
NGC6081           & 0.01679& 13.23 $\pm$  0.01 &  0.94 $\pm$  0.01& S0aA &   0& 10.87 $\pm$  0.20&   \nodata&   \nodata \\
UGC10297          & 0.01757& 14.34 $\pm$  0.02 &  0.60 $\pm$  0.03& ScA  &   5&  9.26 $\pm$  0.16&   8.41 $\pm$   0.09&  -0.35 $\pm$   0.01 \\
UGC10331          & 0.01589& 14.22 $\pm$  0.03 &  0.54 $\pm$  0.03& ScAB &   9&  9.92 $\pm$  0.15&   8.48 $\pm$   0.08&   0.63 $\pm$   0.01 \\
NGC6154           & 0.01993& 13.40 $\pm$  0.01 &  0.82 $\pm$  0.02& SabB &  15& 10.81 $\pm$  0.19&   8.68 $\pm$   0.04&   0.50 $\pm$   0.01 \\
NGC6155           & 0.01802& 12.87 $\pm$  0.01 &  0.64 $\pm$  0.01& ScA  &  32& 10.03 $\pm$  0.16&   8.72 $\pm$   0.05&   0.42 $\pm$   0.01 \\
UGC10384          & 0.01653& 14.25 $\pm$  0.02 &  0.75 $\pm$  0.02& SbA  &   1& 10.21 $\pm$  0.18&   \nodata&   \nodata \\
NGC6168           & 0.01833& 14.27 $\pm$  0.02 &  0.65 $\pm$  0.02& ScAB &  11&  9.46 $\pm$  0.16&   8.51 $\pm$   0.05&   0.14 $\pm$   0.01 \\
NGC6186           & 0.01957& 12.89 $\pm$  0.01 &  0.83 $\pm$  0.01& SbB  &   9& 10.36 $\pm$  0.19&   8.78 $\pm$   0.04&   0.48 $\pm$   0.01 \\
UGC10650          & 0.01097& 15.04 $\pm$  0.03 &  0.58 $\pm$  0.03& ScdA &   5&  9.27 $\pm$  0.16&   8.40 $\pm$   0.02&   0.06 $\pm$   0.01 \\
UGC10710          & 0.02785& 14.05 $\pm$  0.02 &  0.82 $\pm$  0.02& SbA  &   0& 10.87 $\pm$  0.19&   \nodata&   \nodata \\
NGC6310           & 0.01128& 13.19 $\pm$  0.01 &  0.82 $\pm$  0.01& SbA  &   1& 10.39 $\pm$  0.19&   \nodata&   \nodata \\
UGC10796          & 0.01040& 14.45 $\pm$  0.03 &  0.47 $\pm$  0.03& ScdAB&   4&  9.39 $\pm$  0.14&   8.44 $\pm$   0.07&  -0.30 $\pm$   0.02 \\
NGC6361           & 0.01267& 13.06 $\pm$  0.02 &  0.99 $\pm$  0.02& SabA &   6& 10.73 $\pm$  0.21&   8.68 $\pm$   0.04&   0.98 $\pm$   0.01 \\
UGC10811          & 0.02905& 14.16 $\pm$  0.02 &  0.82 $\pm$  0.03& SbB  &   0& 10.89 $\pm$  0.19&   \nodata&   \nodata \\
IC1256            & 0.01568& 13.60 $\pm$  0.01 &  0.72 $\pm$  0.01& SbAB &  25& 10.40 $\pm$  0.17&   8.76 $\pm$   0.05&   0.42 $\pm$   0.01 \\
NGC6394           & 0.02879& 14.14 $\pm$  0.02 &  0.85 $\pm$  0.03& SbcB &   0& 10.89 $\pm$  0.19&   \nodata&   \nodata \\
UGC10905          & 0.02580& 13.31 $\pm$  0.01 &  0.95 $\pm$  0.02& S0aA &   0& 11.25 $\pm$  0.20&   \nodata&   \nodata \\
NGC6478           & 0.02234& 13.76 $\pm$  0.01 &  0.82 $\pm$  0.02& ScA  &  30& 10.79 $\pm$  0.19&   8.67 $\pm$   0.05&   1.06 $\pm$   0.01 \\
NGC6497           & 0.02013& 13.26 $\pm$  0.01 &  0.82 $\pm$  0.02& SabB &   9& 10.91 $\pm$  0.19&   8.70 $\pm$   0.01&   0.61 $\pm$   0.01 \\
UGC11262          & 0.01836& 14.81 $\pm$  0.04 &  0.59 $\pm$  0.05& ScA  &   8&  9.91 $\pm$  0.16&   8.62 $\pm$   0.04&  -0.05 $\pm$   0.01 \\
NGC6978           & 0.01986& 13.15 $\pm$  0.02 &  0.84 $\pm$  0.02& SbAB &   9& 10.94 $\pm$  0.19&   8.73 $\pm$   0.04&   0.60 $\pm$   0.01 \\
UGC11649          & 0.01249& 13.22 $\pm$  0.03 &  0.85 $\pm$  0.03& SabB &  18& 10.50 $\pm$  0.19&   8.69 $\pm$   0.09&  -0.13 $\pm$   0.02 \\
UGC11680NED01     & 0.02649& 13.86 $\pm$  0.02 &  0.90 $\pm$  0.03& SbB  &  10& 11.01 $\pm$  0.20&   8.68 $\pm$   0.06&   0.87 $\pm$   0.01 \\
NGC7047           & 0.01907& 13.41 $\pm$  0.02 &  0.83 $\pm$  0.02& SbcB &   0& 10.82 $\pm$  0.19&   \nodata&   \nodata \\
UGC11717          & 0.02128& 13.90 $\pm$  0.02 &  1.03 $\pm$  0.03& SabA &   2& 10.94 $\pm$  0.21&   \nodata&   \nodata \\
MCG-01-54-016     & 0.01062& 14.83 $\pm$  0.05 &  0.42 $\pm$  0.06& ScdA &   1&  9.17 $\pm$  0.14&   \nodata&   \nodata \\
UGC11740          & 0.02118& 14.26 $\pm$  0.03 &  0.71 $\pm$  0.04& SbcA &   9& 10.42 $\pm$  0.17&   8.65 $\pm$   0.03&   0.57 $\pm$   0.01 \\
UGC11792          & 0.01567& 14.49 $\pm$  0.04 &  0.75 $\pm$  0.05& SbcA &   4& 10.06 $\pm$  0.18&   8.68 $\pm$   0.01&   0.44 $\pm$   0.01 \\
UGC12054          & 0.01674& 14.42 $\pm$  0.03 &  0.51 $\pm$  0.04& ScA  &   5&  8.97 $\pm$  0.15&   8.32 $\pm$   0.05&  -0.59 $\pm$   0.01 \\
NGC7311           & 0.01503& 12.19 $\pm$  0.01 &  0.88 $\pm$  0.01& SaA  &  10& 11.17 $\pm$  0.19&   8.75 $\pm$   0.02&   0.72 $\pm$   0.01 \\
NGC7321           & 0.02362& 13.16 $\pm$  0.01 &  0.75 $\pm$  0.02& SbcB &  37& 11.01 $\pm$  0.18&   8.71 $\pm$   0.05&   0.98 $\pm$   0.01 \\
UGC12185          & 0.02184& 13.82 $\pm$  0.02 &  0.76 $\pm$  0.02& SbB  &   6& 10.66 $\pm$  0.18&   8.66 $\pm$   0.01&   0.29 $\pm$   0.01 \\
UGC12224          & 0.01171& 13.66 $\pm$  0.03 &  0.84 $\pm$  0.04& ScA  &  48& 10.23 $\pm$  0.19&   8.64 $\pm$   0.08&   0.07 $\pm$   0.01 \\
UGC12308          & 0.01724& 14.56 $\pm$  0.04 &  0.45 $\pm$  0.05& ScdA &  10&  8.99 $\pm$  0.14&   8.30 $\pm$   0.08&  -0.52 $\pm$   0.01 \\
NGC7466           & 0.02489& 13.80 $\pm$  0.02 &  0.75 $\pm$  0.02& SbcA &  20& 10.77 $\pm$  0.18&   8.66 $\pm$   0.03&   0.84 $\pm$   0.01 \\
NGC7489           & 0.02072& 13.37 $\pm$  0.02 &  0.55 $\pm$  0.02& SbcA &  24& 10.56 $\pm$  0.15&   8.52 $\pm$   0.13&   1.08 $\pm$   0.01 \\
NGC7536           & 0.01540& 13.42 $\pm$  0.02 &  0.67 $\pm$  0.03& ScAB &  29& 10.42 $\pm$  0.17&   8.61 $\pm$   0.08&   0.70 $\pm$   0.01 \\
NGC7549           & 0.01549& 13.30 $\pm$  0.01 &  0.75 $\pm$  0.02& SbcB &  10& 10.56 $\pm$  0.18&   8.73 $\pm$   0.06&   0.85 $\pm$   0.01 \\
NGC7591           & 0.01633& 13.10 $\pm$  0.01 &  0.84 $\pm$  0.02& SbcB &  27& 10.80 $\pm$  0.19&   8.72 $\pm$   0.05&   1.20 $\pm$   0.01 \\
UGC12494          & 0.01318& 14.58 $\pm$  0.05 &  0.47 $\pm$  0.06& SdB  &   0&  9.56 $\pm$  0.14&   \nodata&   \nodata \\
NGC7608           & 0.01149& 14.03 $\pm$  0.03 &  0.82 $\pm$  0.03& SbcA &   9& 10.06 $\pm$  0.19&   8.73 $\pm$   0.03&   0.08 $\pm$   0.01 \\
UGC12519          & 0.01456& 13.87 $\pm$  0.02 &  0.72 $\pm$  0.02& ScAB &   5& 10.22 $\pm$  0.17&   8.59 $\pm$   0.09&   0.57 $\pm$   0.01 \\
NGC7722           & 0.01327& 12.65 $\pm$  0.01 &  1.00 $\pm$  0.01& SabA &   0& 10.98 $\pm$  0.21&   \nodata&   \nodata \\
UGC12864          & 0.01496& 14.12 $\pm$  0.03 &  0.65 $\pm$  0.03& ScB  &   8& 10.03 $\pm$  0.16&   8.53 $\pm$   0.07&   0.18 $\pm$   0.01 \\
NGC5947           & 0.01974& 13.68 $\pm$  0.01 &  0.72 $\pm$  0.02& SbcB &  35& 10.58 $\pm$  0.17&   8.66 $\pm$   0.07&   0.59 $\pm$   0.01 \\
NGC4676B          & 0.02195& 16.55 $\pm$  0.01 &  1.07 $\pm$  0.02& Sbc  &   2&  9.88 $\pm$  0.22&   \nodata&   \nodata \\
\hline                    
 \end{tabular}
 \end{center}
\end{table*}